\documentclass[12pt]{article}

\usepackage[utf8]{inputenc}
\usepackage{graphicx}
\usepackage{amsmath}
\usepackage{booktabs}
\mathcode`e=`e
\mathcode`d=`d

\DeclareMathOperator{\sign}{sign}
\newtheorem{theorem}{Theorem}
\newtheorem{proposition}{Proposition}

\begin{document}

\title{Bifurcations in economic growth model with distributed time delay transformed to ODE}

\author{Luca Guerrini\\
\footnotesize{Department of Management, Polytechnic University of Marche,} \\
\footnotesize{Piazza Martelli 8, I--60121, Ancona (AN), Italy;} \\
\footnotesize{e-mail: luca.guerrini@univpm.it}\\[0.5cm]
Adam Krawiec \\
\footnotesize{Institute of Economics, Finance and Management, Jagiellonian University,} \\
\footnotesize{\L ojasiewicza 4, 30-348 Krak\'{o}w, Poland;}\\
\footnotesize{e-mail: adam.krawiec@uj.edu.pl}\\[0.5cm]
Marek Szyd{\l}owski \\
\footnotesize{Astronomical Observatory, Jagiellonian University,} \\
\footnotesize{Orla 171, 30-244 Krak\'{o}w, Poland;} \\
\footnotesize{Mark Kac Complex Systems Research Centre, Jagiellonian University,} \\
\footnotesize{\L ojasiewicza 11, 30-348 Krak\'{o}w, Poland;} \\
\footnotesize{email: marek.szydlowski@uj.edu.pl}
}

\date{}

\maketitle

\begin{abstract}
We consider the model of economic growth with time delayed investment function. Assuming the investment is time distributed we can use the linear chain trick technique to transform delay differential equation system to equivalent system of ordinary differential system (ODE). The time delay parameter is a mean time delay of gamma distribution. We reduce the system with distribution delay to both three and four-dimensional ODEs. We study the Hopf bifurcation in these systems with respect to two parameters: the time delay parameter and the rate of growth parameter. We derive the results from the analytical as well as numerical investigations. From the former we obtain the sufficient criteria on the existence and stability of a limit cycle solution through the Hopf bifurcation. 
In numerical studies with the Dana and Malgrange investment function we found two Hopf bifurcations with respect to the rate growth parameter and detect the existence of stable long-period cycles in the economy. We find that depending on the time delay and adjustment speed parameters the range of admissible values of the rate of growth parameter breaks down into three intervals. First we have stable focus, then the limit cycle and again the stable solution with two Hopf bifurcations. Such behaviour appears for some middle interval of admissible range of values of the rate of growth parameter.
\end{abstract}

\section{Introduction}

In economics many processes depends on past events, so it is natural to use the time delay differential equations to model economic phenomena. Two main areas of applications are business cycle and economic growth theories. In last decades the study of impact of investment delay was a subject of an detailed scrutiny as a mechanism for endogenous cycle to describe business cycles and growth cycle. One of the most influential models of business cycle with time delay was the Kaldor-Kalecki model \cite{Krawiec:199989} which was based on the Kaldor model, one earliest endogenous business cycle model \cite{Kaldor:1940mt,Chang:197137}. The Kaldor is a prototype of a dynamical system with cyclic behaviour in which nonlinearity plays a crucial role to generate the endogenous cycles. The nonlinearities are common feature used to model the complexity of economic systems \cite{Chiarella:1990en}. In turn, the investment delay was assumed to be the average time of making investment as it was proposed by Kalecki \cite{Kalecki:1935mt}. Apart from the pioneering interest in time delay in economics, the time delay systems were studied in many domains of science \cite{Lakshmanan:2010dn}.

The Kaldor-Kalecki business cycle model was a subject of many studies as well as augmentations. This model was modified by incorporating the exponential trend to describe growth of an economy \cite{Krawiec:2017175}. This new Kaldor-Kalecki growth model was formulated as similar way as the Kaldor growth model was obtained from the Kaldor business cycle model \cite{Dana:1984dd}.

The models with distributed delays are a more realistic description of economic systems with time delay. There are some examples of such models in context of economic growth \cite{Tarasov20191313}.

In this paper we modify the Kaldor-Kalecki growth model to allow for distributed time of investment. Instead of average time of investment completion the distributed time length of investment is considered. This approach provides more realistic characteristic of investment processes. This growth model has the form of the dynamical system with a distributed time delay. The unimodal distribution function for investment is assumed.

While the delay differential equations methods are developing rapidly, the mathematical methods for ordinary differential equations are superlative, especially when distributed delays are considered. Therefore, it is convenient to approximate a system with distributed delay with a system of ordinary differential equations. One of the method to reduce a system with the distributed delay to an ordinary differential equation system is the linear chain trick \cite{Vogel:1965se,Fargue:1973rs,MacDonald:1978tl}. In consequence infinite-dimensional dynamical systems are approximated by finite-dimensional dynamical system, where the dimension of the system can be chosen.

The main aim of the paper is to study possible bifurcation due to change of the parameter values of the Kaldor-Kalecki growth model. We consider two simplest cases of three and four dimensional dynamical systems obtained through the linear chain trick from the Kaldor-Kalecki growth model with distributed delay. For both models we establish the conditions for existence of the Hopf bifurcation with respect to the time delay parameter and the rate of growth parameter. We show the both parameters play role in a scenario leading to the Hopf bifurcation and arising cyclic behaviour.

In the numerical part of the paper we determine in details the ranges of parameter values for which cyclical behaviour is possible. For these analysis we use the investment function obtained by Dana and Malgrange for French economy \cite{Dana:1984dd}. As in theoretical part of the paper we choose the time delay parameter and the rate of growth parameter. Additionally we consider the adjustment parameter. We show that the combination of these three parameters of model can lead to arising the cycles through Hopf. In the space of these three parameters of the model we obtain the surface (a section of a paraboloid) separating the regions with stable and cyclic solutions.

\section{Model}

In \textit{Economic growth cycles driven by investment delay}, Krawiec and Szyd{\l}owski \cite{Krawiec:2017175} formulated the model based on the Kaldor business cycle model with two modifications: exponential growth introduced by Dana and Malgrange \cite{Dana:1984dd} and Kaleckian investment time delay \cite{Krawiec:199989}. This model of economic growth is described by the following system of differential equations with time delay $\tau \geq 0$,%
\begin{align}
\dot{y}(t) &= \alpha \lbrack I(y(t),k(t))-\gamma
y(t)+G_{0}]-gy(t),
\label{eq:1} \\
\dot{k}(t) &= I(y(t-\tau ),k(t))-(g+\delta )k(t),
\label{eq:2}
\end{align}%
where $I(y(t),k(t))=k(t)\Phi (y(t)/k(t)),$ with
\begin{equation*}
\Phi (y/k)=c+\frac{d}{1+e^{-a\left( vy/k-1\right) }},
\end{equation*}%
and $\alpha ,\gamma ,g,\delta ,G_{0},g$ and $a,c,d,v$ are positive
constants. It can be found that the system has a unique fixed point $%
(y^{*},k^{*})$ with positive coordinates, where%
\begin{equation*}
y^{*}=x^{*}k^{*}\quad \text{and}\quad k^{*}=\frac{\alpha
G_{0}}{gx^{*}+\alpha \left[ sx^{*}-(g+\delta )\right] },
\end{equation*}%
with $x^{*}$ the unique solution of the equation
\begin{equation*}
\Phi (x^{*})=g+\delta .
\end{equation*}%
Because of the S-shape of function $\Phi (x),$ we have that $x^{*}$
always exists and the values of $y^{*}$ and $k^{*}$ depend only on $%
x^{*}$ (in our case, $c<g+\delta <c+d$). Notice that, for economic
considerations, the investment function $I(y(t),k(t))$ is such that
\begin{equation}
I_{y}^{*}=I_{y}(y^{*},k^{*})=\frac{adve^{-a(vx^{*}-1)}}{%
[1+e^{-a(vx^{*}-1)}]^{2}}>0
\label{eq:3}
\end{equation}
and
\begin{equation}
I_{k}^{\ast
}=I_{k}(y^{*},k^{*})=g+\delta -x^{*}I_{y}^{*}<0.
\label{eq:4}
\end{equation}

In this paper, we generalize their model by replacing the time delay
in (\ref{eq:2}) with a distributed delay as follows,
\begin{align}
\dot{y}(t) &= \alpha \lbrack I(y(t),k(t))-\gamma
y(t)+G_{0}]-gy(t),
\label{eq:5} \\
\dot{k}(t) &= I \left(\int\limits_{-\infty }^{t}y(r)\kappa
(t-r)dr,k(t)\right)-(g+\delta )k(t),
\label{eq:6}
\end{align}
where $\kappa (\cdot )$ is a gamma distribution, i.e.%
\begin{equation*}
\kappa (\xi )=\left(\frac{m}{T}\right)^{m}
\dfrac{\xi ^{m-1}e^{-\frac{m}{T}\xi }}{(m-1)!},
\end{equation*}
with $m$ a positive integer that determines the shape of the weighting
function. $T\geq 0$ is a parameter associated with the mean time delay of
the distribution. Notice that as $T\rightarrow 0$ the distribution function
approaches the Dirac distribution, and, thus, one recovers the time delay
case.

Henceforth, we will consider only the cases $m=1$ (weak\ delay kernel) and $%
m=2$ (strong delay kernel). Using the so-called linear chain trick technique
\cite{MacDonald:1978tl},
system (\ref{eq:5})-(\ref{eq:6}) can be transformed into equivalent systems of ODEs. More
precisely, defining the new variable%
\begin{equation*}
u(t)=\int\limits_{-\infty }^{t}y(r)\left( \frac{1}{T}\right) e^{-\frac{1}{T}%
(t-r)}dr
\end{equation*}%
one has the system (case $m=1$)
\begin{align}
\dot{y}(t) & = \alpha \lbrack I(y(t),k(t))-\gamma
y(t)+G_{0}]-gy(t),
\label{eq:7} \\
\dot{u}(t) & = \dfrac{1}{T}\left[ y(t)-u(t)\right], \\
\dot{k}(t) & = I(u(t),k(t))-(g+\delta )k(t),
\label{eq:9}
\end{align}
while defining the new variables
\begin{equation*}
p(t)=\int\limits_{-\infty }^{t}y(r)\left(\frac{2}{T}\right)^{2}(t-r)e^{-\frac{2}{T%
}(t-r)}dr
\end{equation*}
and
\begin{equation*}
w(t)=\int\limits_{-\infty }^{t}y(r)\left(
\frac{2}{T}\right) e^{-\frac{2}{T}(t-r)}dr,
\end{equation*}%
one obtains the system (case $m=2$)
\begin{align}
\dot{y}(t) & = \alpha \lbrack I(y(t),k(t))-\gamma
y(t)+G_{0}]-gy(t),
\label{eq:10} \\
\dot{p}(t) & = \dfrac{2}{T}\left[ w(t)-p(t)\right], \\
\dot{w}(t) & = \dfrac{2}{T}\left[ y(t)-w(t\right] ), \\
\dot{k}(t) & = I(p(t),k(t))-(g+\delta )k(t).
\label{eq:13}
\end{align}

We will now analyse the stability and Hopf bifurcation of systems (\ref{eq:7})-(\ref{eq:9})
and (\ref{eq:10})-(\ref{eq:13}) by determining eigenvalues of linearised systems around the
critical point $(y^{*},y^{*},k^{*})$ and $(y^{*},y^{\ast
},y^{*},k^{*})$, respectively.

\section{The time delay bifurcation analysis}

\section*{Case $m=1$}

The characteristic equation of the linearised system (\ref{eq:7})-(\ref{eq:9}) at the
critical point $(y^{*},u^{*},k^{*}),$ where $u^{*}=y^{\ast
}, $ is given by%
\begin{equation}
\left\vert
\begin{array}{ccc}
\alpha I_{y}^{*}-\alpha \gamma -g-\lambda \quad & \quad 0 & \quad \alpha
I_{k}^{*}\medskip \\
\dfrac{1}{T} & \quad -\dfrac{1}{T}-\lambda & \quad 0\medskip \\
0 & \quad I_{y}^{*} & \quad I_{k}^{*}-(g+\delta )-\lambda%
\end{array}%
\right\vert =0,  \label{eq:14}
\end{equation}%
where $\lambda $ denotes a characteristic root. A direct calculation implies
that (\ref{eq:14}) leads to%
\begin{equation}
\lambda ^{3}+a_{1}(T)\lambda ^{2}+a_{2}(T)\lambda +a_{3}(T)=0,  \label{eq:15}
\end{equation}%
where%
\begin{equation*}
a_{1}(T)=\dfrac{1}{T}-A,\quad \quad a_{2}(T)=-\dfrac{A}{T}-B,\quad \quad
a_{3}(T)=\frac{1}{T}\left( -B-\alpha I_{k}^{*}I_{y}^{*}\right) .
\end{equation*}%
with%
\begin{equation*}
A=\alpha \left( I_{y}^{*}-\gamma \right) -g-x^{*}I_{y}^{*}\quad
\text{and}\quad B=\left[ \alpha \left( I_{y}^{*}-\gamma \right) -g\right]
x^{*}I_{y}^{*}.
\end{equation*}%
The necessary and sufficient condition for the local stability of the
equilibrium point is that all characteristic roots of (\ref{eq:15}) have
negative real parts, which, from the Routh--Hurwitz condition, is equivalent to $%
a_{1}(T)>0,a_{3}(T)>0$ and $a_{1}(T)a_{2}(T)>a_{3}(T).$ Then, $a_{2}(T)>0$
is necessarily satisfied.

Let us examine whether these inequalities hold. First, we notice that $A<0.$
In fact, by contradiction, if $A=0$, then $a_{2}(T)=-\left[ x^{\ast
}I_{y}^{*}\right] ^{2}<0.$ On the other hand, if $A>0$, then $B>0,$ and
so $a_{2}(T)<0.$ The fact $A<0$ implies that $a_{1}(T)>0$ holds always true,
while the inequality $a_{3}(T)>0$ is valid if and only if $B+\alpha
I_{k}^{*}I_{y}^{*}<0.$ Thus, $a_{3}(T)>0$ is always satisfied when $%
B\leq 0,$ and it is verified for $g+\delta -(g+\alpha \gamma )x^{*}<0$
when $B>0.$ Finally, let us consider $a_{1}(T)a_{2}(T)>a_{3}(T).$ Since
\begin{equation*}
a_{1}(T)a_{2}(T)-a_{3}(T)=\frac{(AB)T^{2}+(A^{2}+\alpha I_{k}^{\ast
}I_{y}^{*})T-A}{T^{2}},
\end{equation*}%
the sign of $a_{1}(T)a_{2}(T)-a_{3}(T)$ depends on the sign of $%
(AB)T^{2}+(A^{2}+\alpha I_{k}^{*}I_{y}^{*})T-A,$ which is a
quadratic polynomial in $T.$ We have now several cases.

\begin{enumerate}
\item[i)] If $B=0,$ then $a_{1}(T)a_{2}(T)-a_{3}(T)>0$ holds true if $%
A^{2}+\alpha I_{k}^{*}I_{y}^{*}\geq 0$ or if $A^{2}+\alpha
I_{k}^{*}I_{y}^{*}<0$ and $T<A/(A^{2}+\alpha I_{k}^{\ast
}I_{y}^{*})$. $I)=T_{0}^{*}$.

\item[ii)] If $B>0,$ then $AB<0$ and $-A>0.$ By Descartes' rule of signs
we find that the polynomial $(AB)T^{2}+(A^{2}+\alpha I_{k}^{\ast
}I_{y}^{*})T-A$ has exactly one positive root $T=T_{1}^{*}.$ Hence, $%
a_{1}(T)a_{2}(T)-a_{3}(T)>0$ if $0<T<T_{1}^{*}.$

\item[iii)] If $B<0,$ then $AB>0$ and $-A>0.$ Applying again the Descartes'
rule of signs we see that $(AB)T^{2}+(A^{2}+\alpha I_{k}^{*}I_{y}^{\ast
})T-A$ has (two) sign changes only if $A^{2}+\alpha I_{k}^{*}I_{y}^{\ast
}<0$, meaning that this polynomial may have two positive roots $T_{2}^{\ast
}<T_{3}^{*}$. If this happens, then $a_{1}(T)a_{2}(T)-a_{3}(T)>0$ if $%
0<T<T_{2}^{*}$ and $T>T_{3}^{*}$.
\end{enumerate}

Let $T=T_{*}$ such that $a_{1}(T_{*})a_{2}(T_{*})-a_{3}(T_{\ast
})=0,$ namely $T_{*}=T_{j}^{*}$ $(j=0,1,2,3).$ The curve $T=T_{*}
$ divides the parameter space into stable and unstable parts. Choosing $T$
as a bifurcation parameter, we apply the Hopf bifurcation theorem to
establish the existence of a cyclical movement. This theorem asserts the
existence of the closed orbit, if the characteristic equation (\ref{eq:15}) has
a pair of pure imaginary roots and a non-zero real root, and if the real
part of the imaginary roots is not stationary with respect to the changes of
the parameter $T$. At the critical value $T=T_{*}$, Eq.~(\ref{eq:15})
factors as
\begin{equation*}
\left[ \lambda +a_{1}(T_{*})\right] \left[ \lambda ^{2}+a_{2}(T_{*})%
\right] =0,
\end{equation*}%
so we have the following three roots $\lambda _{1,2}=\pm i\sqrt{%
a_{2}(T_{*})}=\pm i\omega _{*}$ and $\lambda _{3}=-a_{1}(T_{\ast
})<0.$ Next, let us investigate the sign of the real parts of this roots as 
$T$ varies. A differentiation of (\ref{eq:15}) with respect to $T$ yields%
\begin{equation}
\left[ 3\lambda ^{2}+2a_{1}(T)\lambda +a_{2}(T)\right] \frac{d\lambda }{dT}=-%
\left[ a_{1}^{\prime }(T)\lambda ^{2}+a_{2}^{\prime }(T)\lambda
+a_{3}^{\prime }(T)\right] , \label{eq:16}
\end{equation}%
where
\begin{align*}
a_{1}^{\prime }(T) &= -\dfrac{1}{T^{2}}<0,\\
a_{2}^{\prime }(T) &= \dfrac{A}{T^{2}}<0, \\
a_{3}^{\prime }(T) &= -\frac{1}{T^{2}}
\left(-B-\alpha I_{k}^{*}I_{y}^{*}\right) =-\frac{a_{3}(T)}{T}<0.
\end{align*}%
Then, from (\ref{eq:16}), we get%
\begin{equation*}
{Re}\left( \frac{d\lambda }{dT}\right) _{T=T_{*}}=\frac{%
-a_{1}^{\prime }(T_{*})a_{2}(T_{*})-a_{1}(T_{*})a_{2}^{\prime
}(T_{*})+a_{3}^{\prime }(T_{*})}{2\left[ a_{2}(T_{\ast
})+a_{1}^{2}(T_{*})\right] }.
\end{equation*}%
Since
\begin{equation*}
-a_{1}^{\prime }(T_{*})a_{2}(T_{*})-a_{1}(T_{*})a_{2}^{\prime
}(T_{*})+a_{3}^{\prime }(T_{*})=-\dfrac{A}{T_{*}^{3}}\left(
BT_{*}^{2}+1\right) ,
\end{equation*}%
we obtain%
\begin{equation*}
\sign\left[ {Re}\left( \frac{d\lambda }{dT}\right) _{T=T_{*}}\right]
=\sign\left( BT_{*}^{2}+1\right) .
\end{equation*}%
If $B\geq 0,$ we observe that ${Re}\left( d\lambda /dT\right)
_{T=T_{*}}>0$ $($with $T_{*}=T_{0}^{*},T_{1}^{*})$ holds
always true, whether if $B<0,$ then ${Re}\left( d\lambda /dT\right)
_{T=T^{*}}>0$ $($with $T_{*}=T_{2}^{*},T_{3}^{*})$ if $0<T<1/%
\sqrt{-B},$ and ${Re}\left( d\lambda /dT\right) _{T=T^{*}}<0$ if $%
T>1/\sqrt{-B}.$

The previous analysis leads to the following conclusions.

\begin{theorem}
Let $A<0,$ with $A$ defined as in $(\ref{eq:15})$.

\begin{enumerate}
\item[$1)$] If $B=0$ and $A^{2}+\alpha I_{k}^{*}I_{y}^{*}<0$ or if $%
B>0$ and $B+\alpha I_{k}^{*}I_{y}^{*}<0,$ then there exists $%
T=T_{*}>0$ such that the equilibrium point $(y^{*},y^{*},k^{\ast
})$ of $(\ref{eq:7})$-$(\ref{eq:9})$ is locally asymptotically stable for all $T<T_{*}$
and unstable for $T>T_{*}.$ System $(\ref{eq:7})$-$(\ref{eq:9})$ undergoes a Hopf
bifurcation at $(y^{*},y^{*},k^{*})$ when $T=T_{*}$.

\item[$2)$] If $B<0$, then there exists $0<T_{2}^{*}<T_{3}^{*}$ such
that the equilibrium point $(y^{*},y^{*},k^{*})$ of $(\ref{eq:7})$-$(\ref{eq:9})$
is locally asymptotically stable for all $T<T_{2}^{*}$ and $%
T<T_{3}^{*}$, and unstable for all $T_{2}^{*}<T<T_{3}^{*}.$ A
comparison of $1/\sqrt{-B}$ with $T_{2}^{*}$ and $T_{3}^{*}$ yields
that system $(\ref{eq:7})$-$(\ref{eq:9})$ undergoes a Hopf bifurcation at $(y^{*},y^{\ast
},k^{*})$ when $T=T_{2}^{*}$ or $T=T_{3}^{*}$ or $T=T_{2}^{\ast
} $ and $T=T_{3}^{*}.$
\end{enumerate}
\end{theorem}

\begin{figure}[ht]
\begin{center}
\includegraphics[width=0.8\textwidth]{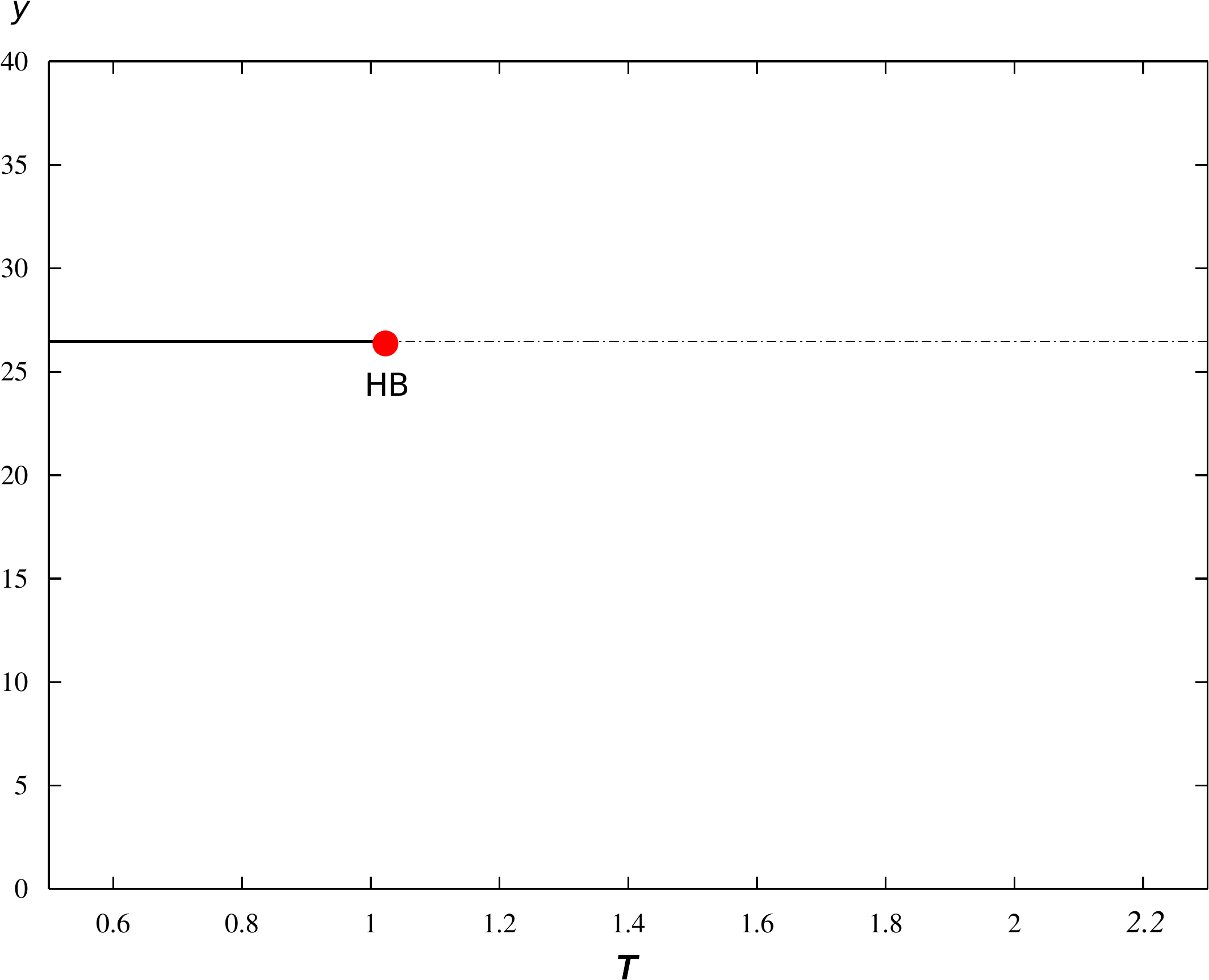}
\end{center}
\caption{\label{fig:1}
The bifurcation diagram for model for system (\ref{eq:7})-(\ref{eq:9}) $m=1$ with investment function (\ref{eq:23}) for delay parameter $T$. The solid line indicates a critical point with asymptotic stability and the dot-dash line corresponds to an unstable critical point with a limit cycle around it.}
\end{figure}

For system (\ref{eq:7})-(\ref{eq:9}) figure~\ref{fig:1} presents the bifurcation diagram for the time delay parameter $T$.

\section*{Case $m=2$}

The characteristic equation of the linearised system (\ref{eq:10})-(\ref{eq:13}) at the
critical point $(y^{*},p^{*},w^{*},k^{*}),$ where $p^{\ast
}=w^{*}=y^{*},$ takes the form%
\begin{equation}
\left\vert
\begin{array}{cccc}
\alpha I_{y}^{*}-\alpha \gamma -g-\lambda \quad & \quad 0\quad & \quad 0
& \quad \alpha I_{k}^{*}\medskip \\
0\quad & -\dfrac{2}{T}-\lambda \quad & \quad \dfrac{2}{T} & \quad 0\medskip
\\
\dfrac{2}{T}\quad & \quad 0\quad & -\dfrac{2}{T}-\lambda & \quad 0\medskip
\\
0\quad & \quad I_{y}^{*}\quad & \quad 0 & \quad I_{k}^{*}-(g+\delta
)-\lambda%
\end{array}%
\right\vert ,  \label{eq:17}
\end{equation}%
where $I_{y}^{*}$ and $I_{k}^{*}$ are defined as in (\ref{eq:3}) and (\ref{eq:4}),
which leads to the following fourth order algebraic equation in $\lambda $
\begin{equation}
\lambda ^{4}+a_{1}(T)\lambda ^{3}+a_{2}(T)\lambda ^{2}+a_{3}(T)\lambda
+a_{4}(T)=0,  \label{eq:18}
\end{equation}%
where%
\begin{equation*}
a_{1}(T)=\dfrac{4}{T}-(M+N),\quad \quad \quad a_{2}(T)=\dfrac{4}{T^{2}}-%
\dfrac{4(M+N)}{T}+MN,
\end{equation*}%
and%
\begin{equation*}
a_{3}(T)=\dfrac{4}{T}\left[ MN-\dfrac{M+N}{T}\right] ,\quad \quad \quad
a_{4}(T)=\dfrac{4(MN+P)}{T^{2}},
\end{equation*}%
with%
\begin{equation}
M=\alpha \left( I_{y}^{*}-\gamma \right) -g,\quad \quad \quad
N=I_{k}^{*}-(g+\delta )<0,\quad \quad \quad P=-\alpha I_{k}^{\ast
}I_{y}^{*}>0.  \label{eq:19}
\end{equation}%
According to the Routh--Hurwitz conditions for stable roots, the equilibrium
point $(y^{*},y^{*},y^{*},k^{*})$ of system (\ref{eq:14}) is locally
asymptotically stable if $a_{1}(T)>0,a_{3}(T)>0,a_{4}(T)>0$ and $%
a_{1}(T)a_{2}(T)a_{3}(T)>a_{3}^{2}(T)+a_{1}^{2}(T)a_{4}(T),$ namely if%
\begin{equation*}
\dfrac{4}{T}-(M+N)>0,\quad \quad \quad MN-\dfrac{M+N}{T}>0,\quad \quad \quad
MN+P>0  \label{10}
\end{equation*}%
and
\begin{multline*}
\varphi (T)=\left[ (M+N)(MN)^{2}\right] T^{4}+[(M+N)^{2}(P-4MN)]T^{3} \\%
[0.2cm]
+\left\{ 4(M+N)\left[ (M+N)^{2}+2MN-2P\right] \right\} T^{2} \\[0.2cm]
+\left\{ 16\left[ P-(M+N)^{2}\right] \right\} T+16(M+N)<0.
\end{multline*}%
Taking in mind that $N<0$ and $P>0,$ we derive that conditions (\ref{10})
hold always true if $M\leq 0.$ On the other hand, when $M>0,$ they are valid
if $M+N<0,$ $MN+P>0$ and $T<(M+N)(/(MN).$ The condition $\varphi (T)<0$ is
difficult to handle, unless $M=0.$ In fact, in this case,
\begin{equation*}
\varphi (T)=\left( N^{2}P\right) T^{3}+\left[ 4N\left( N^{2}-2P\right) %
\right] T^{2}+\left[ 16\left( P-N^{2}\right) \right] T+16N
\end{equation*}
is such that $\varphi (0)<0$ and $\varphi (+\infty )=+\infty .$ Hence, there
exists at least a positive value of $T,$ say $T_{0}^{*},$ such that $%
\varphi (T)=0.$ We need now to recall Descartes' rule of signs and its
corollary, that state \textquotedblleft the number of positive roots of the
polynomial $\varphi (T)$ is either equal to the number of sign differences
between consecutive nonzero coefficients, or is less than it by an even
number\textquotedblright\ and \textquotedblleft the number of negative roots
is the number of sign changes after multiplying the coefficients of
odd-power terms by $-1$, or fewer than it by an even
number\textquotedblright , respectively. Applying these rules to the
polynomial $\varphi (T),$ we get that $\varphi (T)$ has one positive zero
and the number of negative zeros must be either $2$ or $0.$ Therefore, $%
\varphi (T)<0$ if $T<T_{0}^{*}.$ When $M=0,$ the equilibrium point $%
(y^{*},y^{*},y^{*},k^{*})$ of (\ref{eq:14}) is locally
asymptotically stable for $T<T_{0}^{*}$.

Assume there exists $T_{*}>0$ such that $\varphi (T_{*})=0,$ i.e.
\begin{equation*}
a_{1}(T_{*})a_{2}(T_{*})a_{3}(T_{*})-a_{3}^{2}(T_{\ast
})-a_{1}^{2}(T_{*})a_{4}(T_{*})=0.
\end{equation*}%
In this case, we can rewrite the characteristic equation (\ref{eq:18}) as
\begin{equation*}
\left[ a_{1}(T_{*})\lambda ^{2}+a_{3}(T_{*})\right] \left[
a_{1}(T_{*})\lambda ^{2}+a_{1}^{2}(T_{*})\lambda +a_{1}(T_{\ast
})a_{2}(T_{*})-a_{3}(T_{*})\right] =0.
\end{equation*}%
so that we have  two purely imaginary roots
\begin{equation*}
\lambda _{1,2}=\pm i\sqrt{\frac{a_{3}(T_{*})}{a_{1}(T_{*})}}=\pm
i\omega _{*},
\end{equation*}%
and two other roots,
\begin{equation*}
\lambda _{3,4}=\frac{-a_{1}^{2}(T_{*})\pm \sqrt{a_{1}^{4}(T_{\ast
})-4a_{1}(T_{*})\left[ a_{1}(T_{*})a_{2}(T_{*})-a_{3}(T_{*})%
\right] }}{2a_{1}(T_{*})},
\end{equation*}%
which have real parts different from zero since
\begin{equation*}
\lambda _{3}+\lambda _{4}=-a_{1}(T_{*})<0
\end{equation*}
and
\begin{equation*}
\lambda _{3}\lambda _{4}=\left[ a_{1}(T_{*})a_{2}(T_{\ast
})-a_{3}(T_{*})\right] /a_{1}(T_{*})>0.
\end{equation*}%
Differentiating the characteristic equation (\ref{eq:18}) with respect to $T$,
we have%
\begin{equation}
\left[ 4\lambda ^{3}+3a_{1}(T)\lambda ^{2}+2a_{2}(T)\lambda +a_{3}(T)\right]
\frac{d\lambda }{dT}=-\left[ a_{1}^{\prime }(T)\lambda ^{3}+a_{2}^{\prime
}(T)\lambda ^{2}+a_{3}^{\prime }(T)\lambda +a_{4}^{\prime }(T)\right] ,
\label{12}
\end{equation}%
i.e.%
\begin{equation}
\frac{d\lambda }{dT}=-\frac{a_{1}^{\prime }(T)\lambda ^{3}+a_{2}^{\prime
}(T)\lambda ^{2}+a_{3}^{\prime }(T)\lambda +a_{4}^{\prime }(T)}{4\lambda
^{3}+3a_{1}(T)\lambda ^{2}+2a_{2}(T)\lambda +a_{3}(T)},  \label{12bis}
\end{equation}%
where%
\begin{equation*}
a_{1}^{\prime }(T)=-\dfrac{4}{T^{2}},\qquad \quad 
a_{2}^{\prime }(T)=-\dfrac{8}{T^{3}}+\dfrac{4(M+N)}{T^{2}},
\end{equation*}%
and%
\begin{equation*}
a_{3}^{\prime }(T)=-\dfrac{4MN}{T^{2}}+\dfrac{8(M+N)}{T^{3}},\qquad \quad 
a_{4}^{\prime }(T)=-\dfrac{8(MN+P)}{T^{3}}.
\end{equation*}%
Letting $\lambda =i\omega _{*}$ in (\ref{12bis}), a direct calculation
yields%
\begin{equation*}
{Re}\left( \frac{d\lambda }{dT}\right) _{T=T_{*}}=-\frac{%
a_{1}(T_{*})\varphi ^{\prime }(T_{*})}{2\left\{ a_{1}^{3}(T_{\ast
})a_{3}(T_{*})+\left[ a_{1}(T_{*})a_{2}(T_{*})-2a_{3}(T_{*})%
\right] ^{2}\right\} },
\end{equation*}%
where
\begin{multline*}
\varphi ^{\prime }(T_{*})=a_{1}^{\prime }(T_{*})a_{2}(T_{\ast
})a_{3}(T_{*})+a_{1}(T_{*})a_{2}^{\prime }(T_{*})a_{3}(T_{\ast
})+a_{1}(T_{*})a_{2}(T_{*})a_{3}^{\prime }(T_{*}) \\[0.25cm]
-2a_{3}(T_{*})a_{3}^{\prime }(T_{*})-2a_{1}(T_{*})a_{1}^{\prime
}(T_{*})a_{4}(T_{*})-a_{1}^{2}(T_{*})a_{4}^{\prime }(T_{*}).
\end{multline*}%
Let us notice that $\sign\left[ \text{Re}\left( d\lambda /dT\right) _{T=T_{*}}%
\right] =\sign\left[ -\varphi ^{\prime }(T_{*})\right]$, and recall that
$\sign\left[ \text{Re}\left( d\lambda /dT\right) _{T=T_{*}}\right] >0$
and $\sign\left[ \text{Re}\left( d\lambda /dT\right) _{T=T_{*}}\right] <0$
correspond to crossings of the imaginary axis from right to left, and from
left to right, respectively.

Summarizing all the previous analysis, we have the following results.

\begin{theorem}
Let $M$  be defined as in $(\ref{eq:19})$.

\begin{enumerate}
\item[$1)$] Let $M=0.$ There exists $T_{0}^{*}>0$ such that the
equilibrium point $(y^{*},y^{*},y^{*},k^{*})$ of $(\ref{eq:14})$
is locally asymptotically stable for $T<T_{0}^{*}$, unstable for $%
T>T_{0}^{*}$, and bifurcates to a limit cycle through a Hopf bifurcation
at the equilibrium point when $T=T_{0}^{*}$.

\item[$2)$] Let $M\neq 0.$ The equilibrium point $(y^{*},y^{\ast
},y^{*},k^{*})$ of $(\ref{eq:14}$) is locally asymptotically stable if $%
M<0$ and $\varphi (T)<0$ or if $M>0,M+N<0,$ $MN+P>0,T<(M+N)(/(MN)$ and $%
\varphi (T)<0.$ If there exists $T=T_{*}$ such that $\varphi (T_{\ast
})=0$ and $\varphi ^{\prime }(T_{*})\neq 0,$ then a Hopf bifurcation may
occurs at the equilibrium point as $T$ passes through $T_{*}.$
\end{enumerate}
\end{theorem}

\section{The rate of growth bifurcation analysis}


Let us consider the dynamics of the system (\ref{eq:7})-(\ref{eq:9}) with respect to the change of the parameter $g$ (the rate of economic growth).

\begin{proposition}
The critical point of system (\ref{eq:7})-(\ref{eq:9}) (and equivalently system (\ref{eq:1})-(\ref{eq:2}) always exists for the rate of growth parameter $g$ in the interval $c - \delta < g < c + d - \delta$.
\end{proposition}
It is shown earlier that the system (\ref{eq:1})-(\ref{eq:2}) has unique fixed point for $c < g + \delta < c +d$. The economy with the investment function $I(y,k)= k\Phi(y,k)$ has a fixed point or a limit cycle solution only for some rates of growth within the interval $(c - \delta , c + d - \delta)$. The parameters $c$ and $d$ of the investment function (and also capital stock depreciation) put some limit on minimal and maximal rates of growth.

Let us consider the characteristic equation of the linearised system (\ref{eq:10})-(\ref{eq:13}) at the critical point $(y^*, u^*, k^*)$, in the form
\begin{equation} \label{eq:20}
\left| \begin{array}{ccc}
\alpha(I_y^*(g) - \gamma) - g - \lambda & 0 & \alpha (g + \delta - x^*(g) I_y^*(g)) \\ 
\frac{1}{T} & - \frac{1}{T} - \lambda & 0 \\
0 & I_y^* & - x^*(g) I_y^*(g) - \lambda
\end{array} \right| = 0
\end{equation}
or
\begin{equation} \label{eq:21}
\lambda^3 + a_1(g) \lambda^2 + a_2(g) \lambda + a_3(g) = 0
\end{equation}
where
\begin{align*}
a_1(g) &= \frac{1}{T} - \alpha(I_y^*(g) - \gamma) + g + x^*(g) I_y^*(g) \\
a_2(g) &= -\frac{1}{T} [\alpha(I_y^*(g) - \gamma) - g - x^*(g) I_y^*(g)] - [\alpha(I_y^*(g) - \gamma) - g)x^*(g) I_y^*(g) \\
a_3(g) &= \frac{1}{T} [ -[\alpha(I_y^*(g) - \gamma) - g]x^*(g) I_y^*(g) -\alpha I_k^*(g) I_y^*(g)]
\end{align*}
where $\lambda$ is a root of the characteristic equation.

The discriminant of the characteristic equation is 
\begin{equation} \label{eq:22}
\Delta = 18a_1 a_2 a_3 - 4a_2^3 a_3 + a_2^2 a_3^2 -4a_3^3 -27 a_3^2.
\end{equation}
\begin{proposition}
If the expression (\ref{eq:22}) is positive than the all eigenvalues are real and if it is negative there is one real, one pair of conjugate complex eigenvalues. For the zero value of the expression (\ref{eq:22}) the critical point is non-hyperbolic.
\end{proposition}

For real eigenvalues we have the following proposition
\begin{proposition}
In the interval $c - \delta < g < c + d - \delta$, there are two subintervals with the positive values of the discriminant (\ref{eq:22}) there two cases for the values of rate of growth parameter $g$. In these subintervals there are three negative real eigenvalues.
\end{proposition}

The subintervals of the parameter $g$ with two negative and one positive eigenvalues are non-physical regions as the critical point ($y^*, u^*, k^*$) does not lie in a positive quadrant. 

For complex eigenvalues we have the following proposition
\begin{proposition}
In the interval $c - \delta < g < c + d - \delta$ and negative values of the discriminant (\ref{eq:22}) for the increasing value of the rate of growth parameter $g$ there are two supercritical Hopf bifurcations. For the value $g = g_{1,\text{Hopf}}$ the limit cycle is created, and for the value $g=g_{2,\text{Hopf}}$ the limit cycle is destroyed ($g_{1,\text{Hopf}}<g_{2,\text{Hopf}}$).
\end{proposition}

Therefore, as the the rate of growth parameter is increasing in the interval $g_\text{min} = c - \delta < g < c + d - \delta = g_\text{max}$ the eigenvalues change as follows. In the first subinterval $(g_{\text{min}}; g_1)$ there are three real eigenvalues (two negative, one positive). In the second subinterval $(g_1; g_{1,\text{Hopf}})$ there are three real eigenvalues (three negative). In the third subinterval $( g_{1,\text{Hopf}};g_{2,\text{Hopf}})$ there are one real eigenvalue (negative) and one conjugate complex eigenvalue (positive real parts). In the fourth subinterval $(g_{2,\text{Hopf}};g_2)$ there are three real eigenvalues (three negative). And finally, in the fifth subinterval $g_2,(g_{\text{max}})$ there are three real eigenvalues (two negative, one positive).

We some example values of parameters we can determine the values of the rate of growth parameter for which the eigenvalues change their character or sign. We assume the values of investment function parameters obtained by Dana and Malgrange, namely, $c=0.01$, $d=0.026$, $a=9$, $v=4.23$. We fix also the following model parameters $\alpha=1$, $\gamma=0.15$, $\delta=0.007$, $G_0=2$ and $T=1$. The rate of growth parameter $g$ is taken within the interval $g_\text{min} = c - \delta < g < c + d - \delta = g_\text{max}$. The results are presented in table~\ref{tab:3}.

\begin{table}[ht]
\begin{center}
\caption{The intervals of values of rate of growth parameter $g$ and respective signs of eigenvalues of the characteristic equation (\ref{eq:20}). It is assumed that $c=0.01$, $d=0.026$, $a=9$, $v=4.23$ (the investment function), $\alpha=1$, $\gamma=0.15$, $\delta=0.007$, $G_0=2$ and $T=1$ (rest model parameters).\label{tab:3}}
\begin{tabular}{llc}
\toprule
real eigenvalues & complex eigenvalues& rate of growth parameter \\
\midrule
1 negative & pair with negative real part & (0.003, 0.0101198)\\
1 negative & pair with positive real part & (0.0101199, 0.0203258)\\
1 negative & pair with negative real part & (0.0203259, 0.029)\\
\bottomrule
\end{tabular}
\end{center}
\end{table}

\begin{figure}[ht]
\begin{center}
\includegraphics[width=0.8\textwidth]{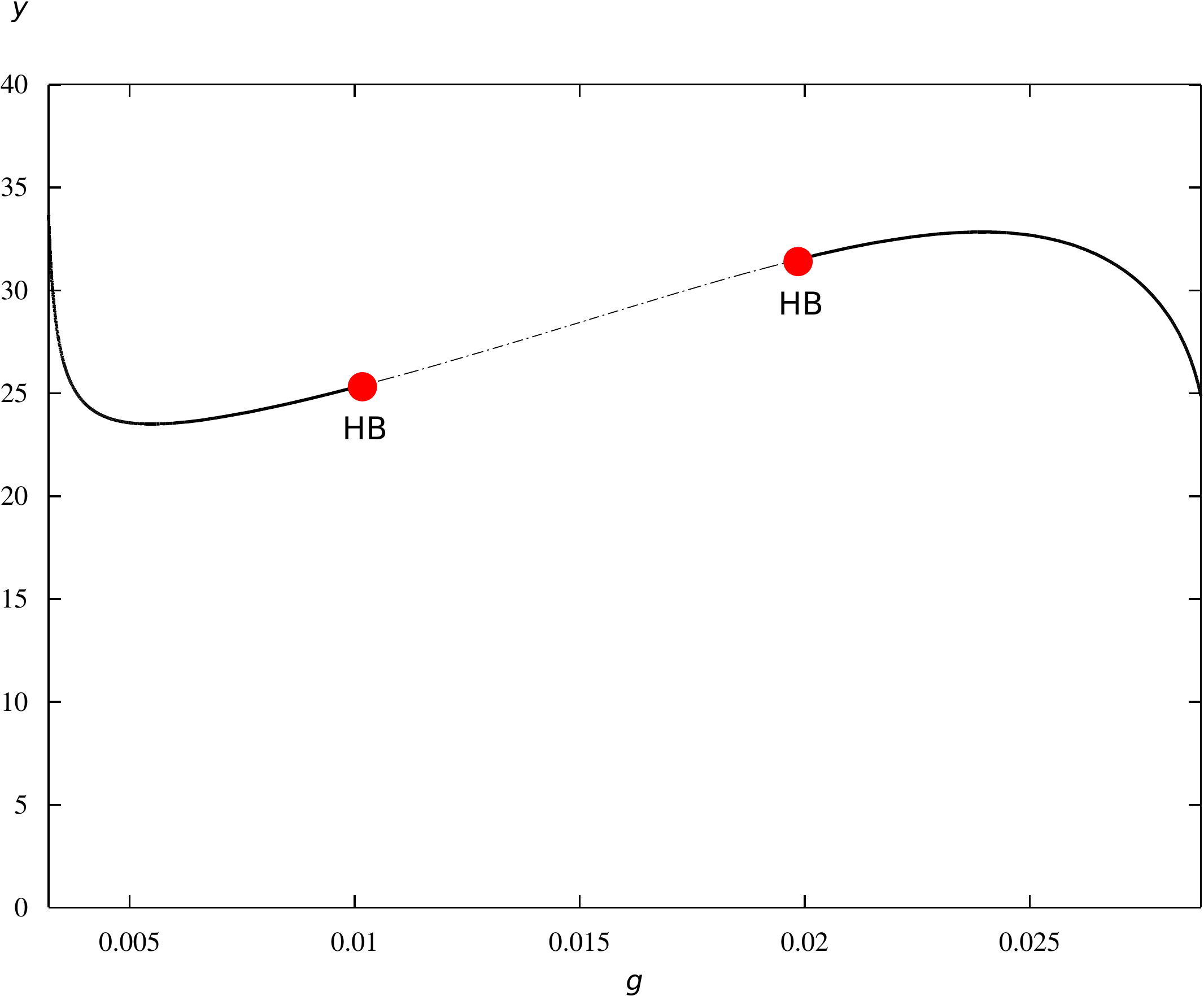}
\end{center}
\caption{\label{fig:2}
The bifurcation diagram for model for system (\ref{eq:7})-(\ref{eq:9}) $m=1$ with investment function (\ref{eq:23}) for delay parameter $T$. The solid line indicates critical point with asymptotic stability and the dot-dash line corresponds to the unstable critical with a limit cycle around it.}
\end{figure}

For system (\ref{eq:7})-(\ref{eq:9}) figure~\ref{fig:2} presents the bifurcation diagram for the rate of growth parameter $g$.

\section{Numerical analysis of the Hopf bifurcation}

The original Kaldor model exhibited the limit cycle behaviour due to the Hopf bifurcation caused by the increase of the parameter $\alpha$ value \cite{Chang:197137}. Later, it has been augmented both by introducing the investment lag $T$ and exogenous growth trend $g$. The increase of the investment time delay parameter value also generates the limit cycle \cite{Krawiec:199989}. However, the dependence the Hopf bifurcation on the rate of growth parameter was not elaborated so far. Both Chang and Smyth in the Kaldor model \cite{Chang:197137} as well as Dana and Malgrange in the Kaldor model with exogenous growth trend investigated the parameter $\alpha$ as the bifurcation parameter\cite{Dana:1984dd}.

We conduct the numerical analysis of both models: $m=1$ and $m=2$. We assume the investment function with the following parameter values: $c=0.01$, $d=0.026$, $a=9$ and $v=4.23$~\cite{Dana:1984dd}
\begin{equation} \label{eq:23}
I(y,k) = k \Phi(y,k) = 0.01 + \frac{0.026}{1+e^{-9(4.23y/k -1)}}.
\end{equation}

We also assume the following the model parameters: $\gamma=0.15$, $\delta=0.007$, $G_0=2$ and consider the three parameters in the following intervals: $T \in (0,5)$, $\alpha \in (0.5, 1.0)$ and $g \in (0.01, 0.02)$.

\section*{Case $m=1$}

In this case we consider the three-dimensional model (\ref{eq:7})-(\ref{eq:9}) for the state variables $(y, u, k)$. In this model we study numerically the stability of the critical point $(y^* = u^*, k^*)$ to find the values of parameter $T$ for which the critical point loses the stability and the limit cycle is created through the Hopf bifurcation mechanism. We study in details the dependence of the bifurcation value of $T$ on the model parameters $\alpha$ and $g$ as well as the dependence the bifurcation value of $g$ on the model parameter $\alpha$ and $T$.

\begin{figure}[ht]
\begin{center}
\includegraphics[width=0.9\textwidth]{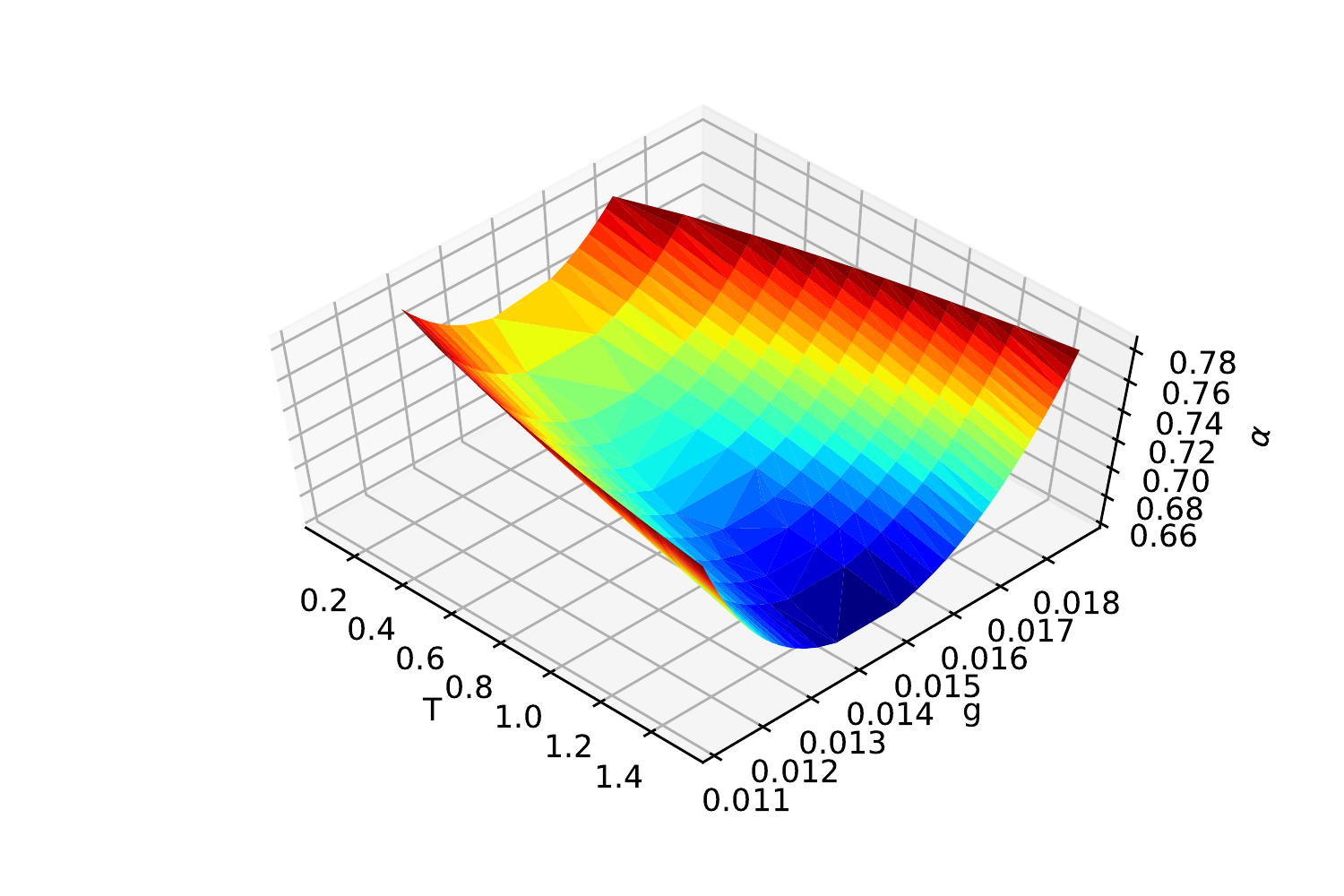}
\end{center}
\caption{\label{fig:3}
The Hopf bifurcation surface in the space of parameters $(\alpha, g, T)$ for system $m=1$ and with investment function (\ref{eq:23}). Outside of the surface is the region of asymptotic stability, while inside of the surface is the region of parameters values for which a limit cycle solution exists.}
\end{figure}

The bifurcation surface in the parameter space $(a,g,T)$ is presented in figure~\ref{fig:3}. The region below the surface corresponds to the asymptotic stability of the critical point $(y^*, u^*, k^*)$. The region inside corresponds to parameter values for which system (4) has an unstable critical point with a limit cycle around it.

\begin{figure}[ht]
\begin{center}
\includegraphics[width=0.49\textwidth]{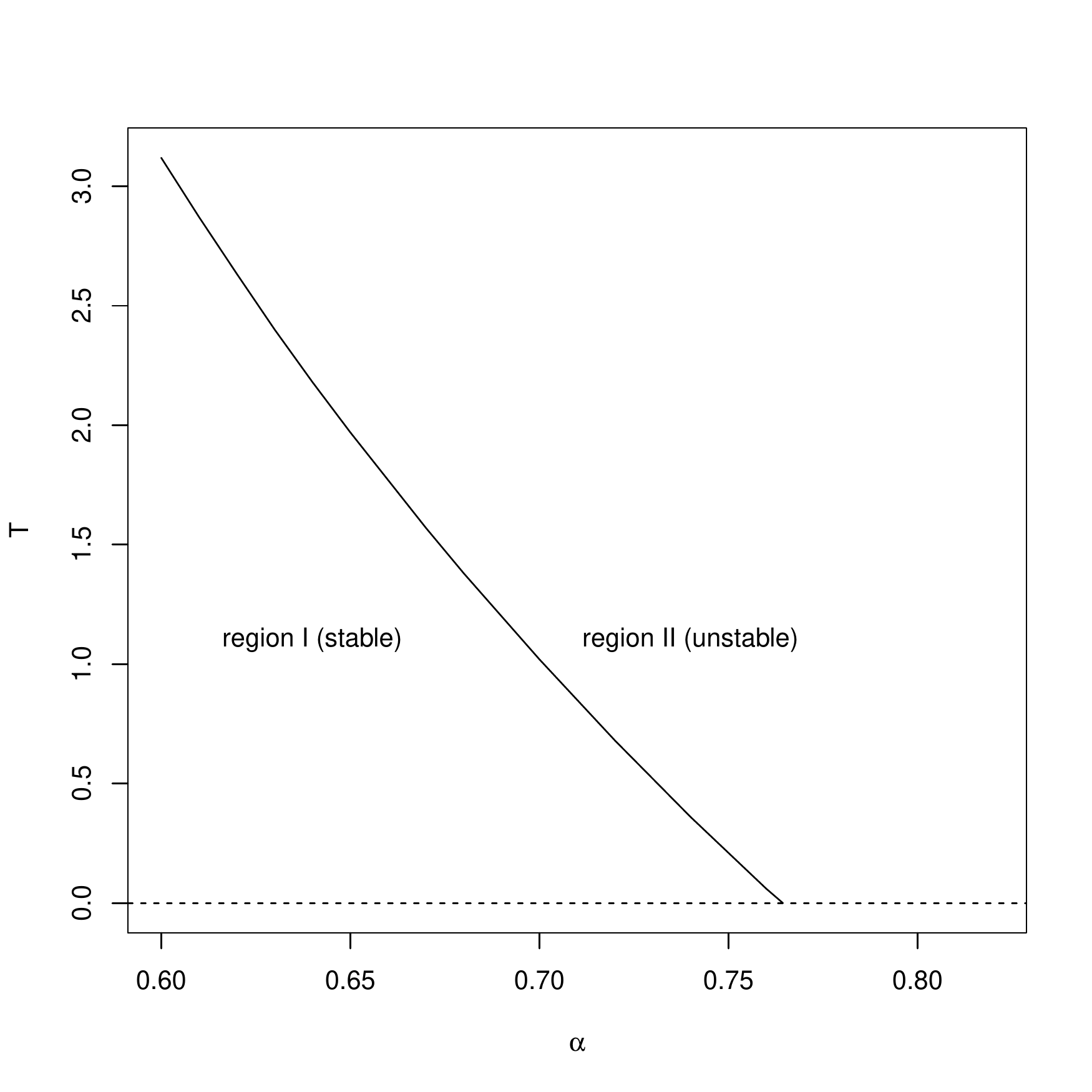}
\end{center}
\caption{\label{fig:4}
The plane of parameters $(\alpha,T)$ for system $m=1$ and $g=0.016$ with investment function (\ref{eq:23}). Region I is the region of asymptotic stability, while region II is the region of parameters value for which a limit cycle solution exists.}
\end{figure}

Let conduct more detailed analysis and consider relations between two parameters with a third parameter fixed. First, the relation of $T$ on $\alpha$ for $g=0.016$ is shown in Fig.~\ref{fig:4}. We find that the asymptotic stability region exist only if $\alpha < 0.7644$ (with $g=0.016$). In the interval of $\alpha \in (0.6, 0.764)$ we find the relation $T_\text{bi}(\alpha)$ is
\begin{equation}
T_{\text{bi}} = -11.137983 + \frac{8.512805}{\alpha}.
\end{equation}

\begin{figure}[ht]
\begin{center}
\includegraphics[width=0.47\textwidth]{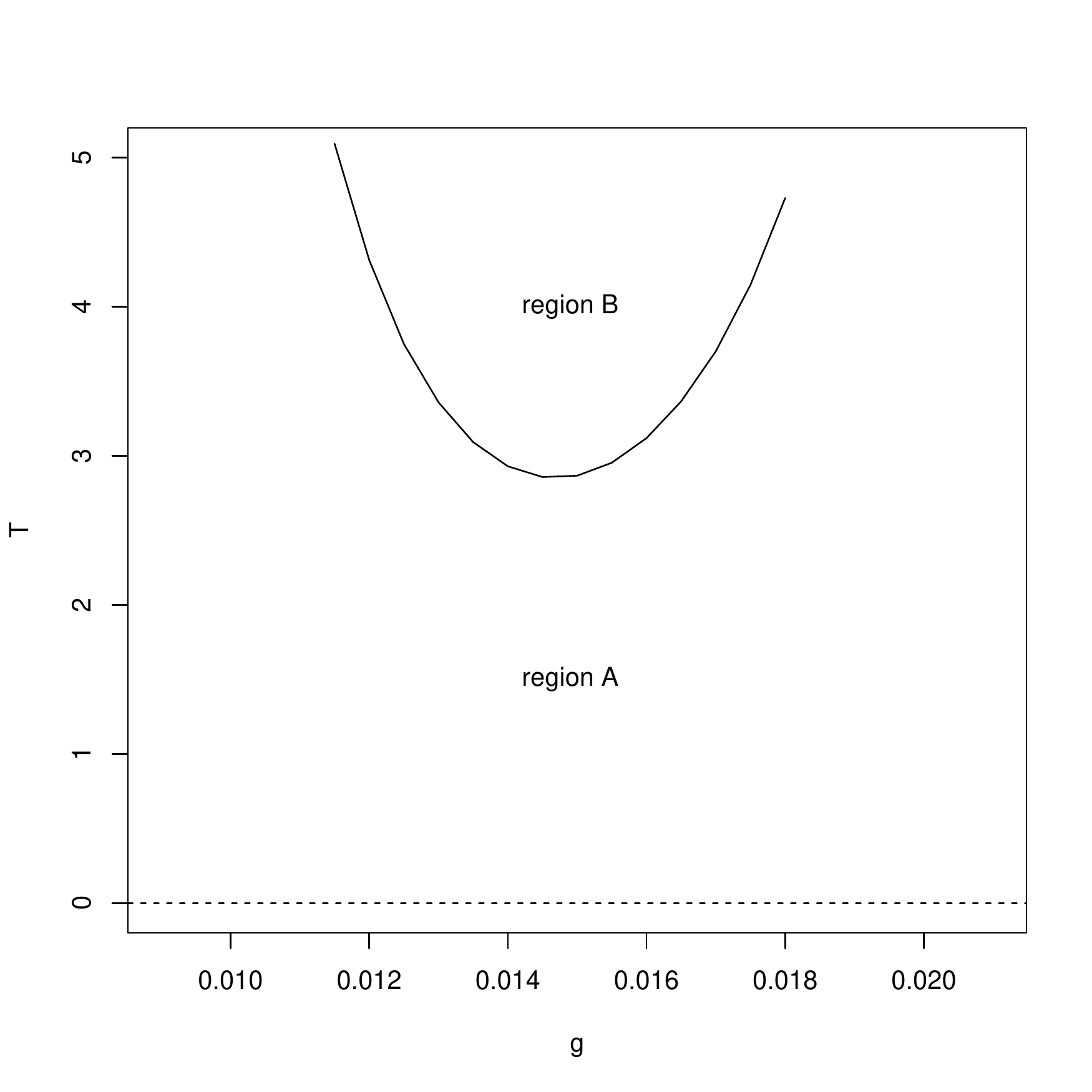} 
\includegraphics[width=0.47\textwidth]{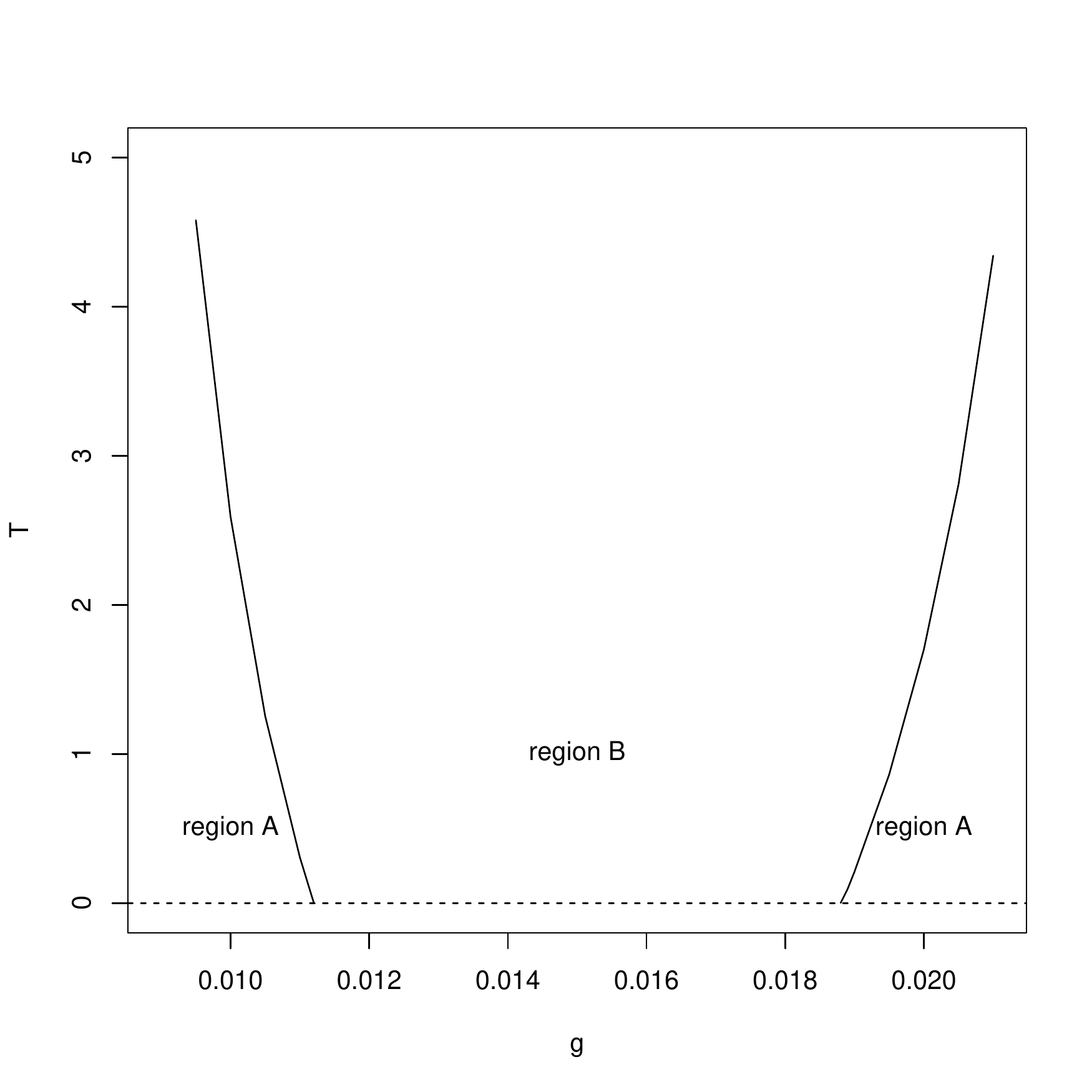}
\end{center}
\caption{\label{fig:5}
The plane of parameters $(g,T)$ for system $m=1$ with investment function (\ref{eq:23}). Here, it is assumed that $\alpha=0.6$ (left panel) and $\alpha = 0.9$ (right panel). Region A is the region of asymptotic stability and region B is the region of limit cycle solution and their are separated by the bifurcation line $T_{\text{bi}}(g)$.}
\end{figure}

Second, we analyze the dependence of the parameter $T$ on the parameter $g$ with the fixed value of $\alpha$. We consider the three values of parameter $\alpha$. The stability regions on the plane $(g,T)$ are shown for $\alpha = 0.6$ and $\alpha = 0.9$ in fig.~\ref{fig:5}. As the value of the parameter $g$ increases the region B is growing.

The bifurcation line separating the region A and B is described by a quadratic equation
\begin{equation}
T_{\text{bi}} = a_2 g^2 + a_1 g + a_0.
\end{equation}

\begin{figure}[ht]
\begin{center}
\includegraphics[width=0.48\textwidth]{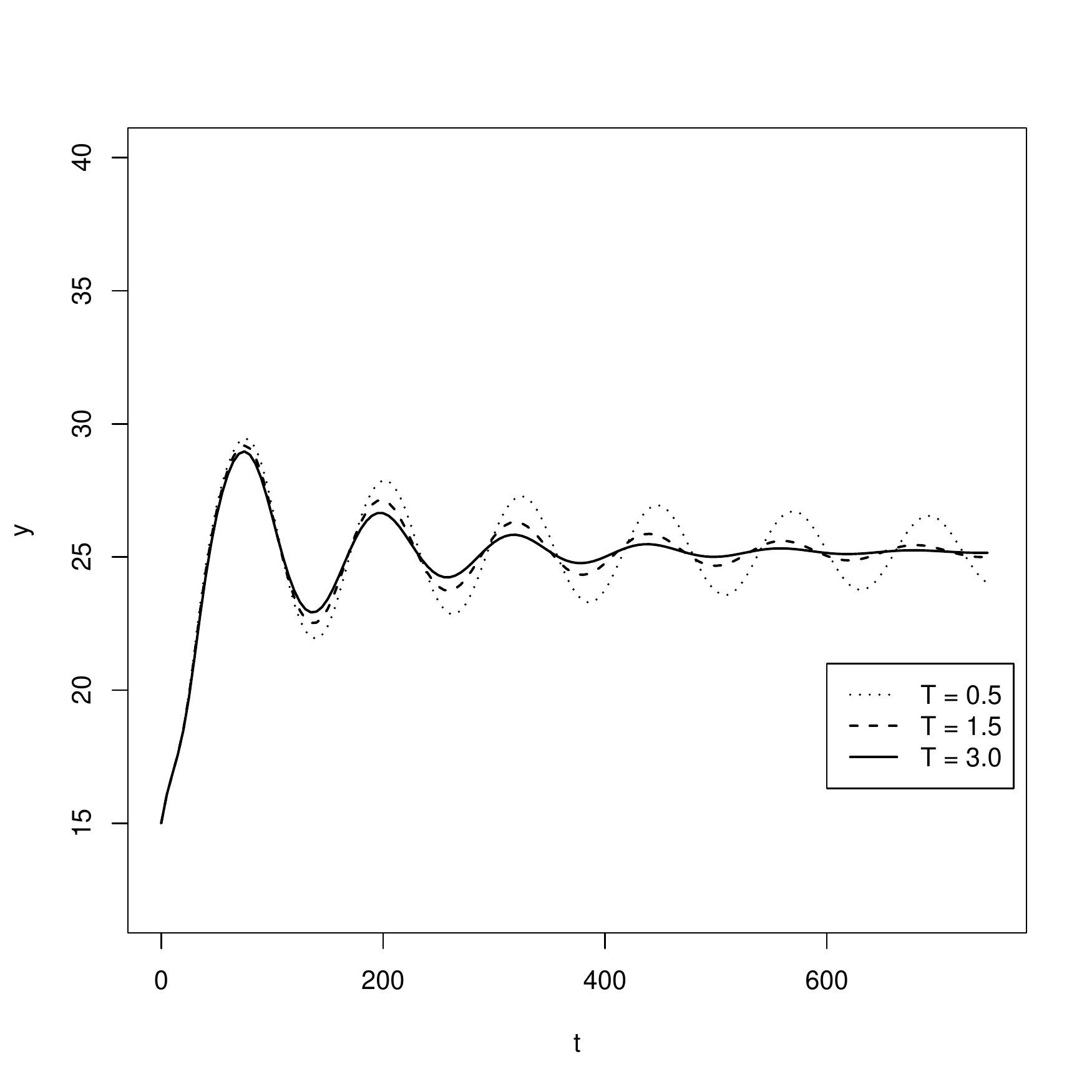}
\includegraphics[width=0.48\textwidth]{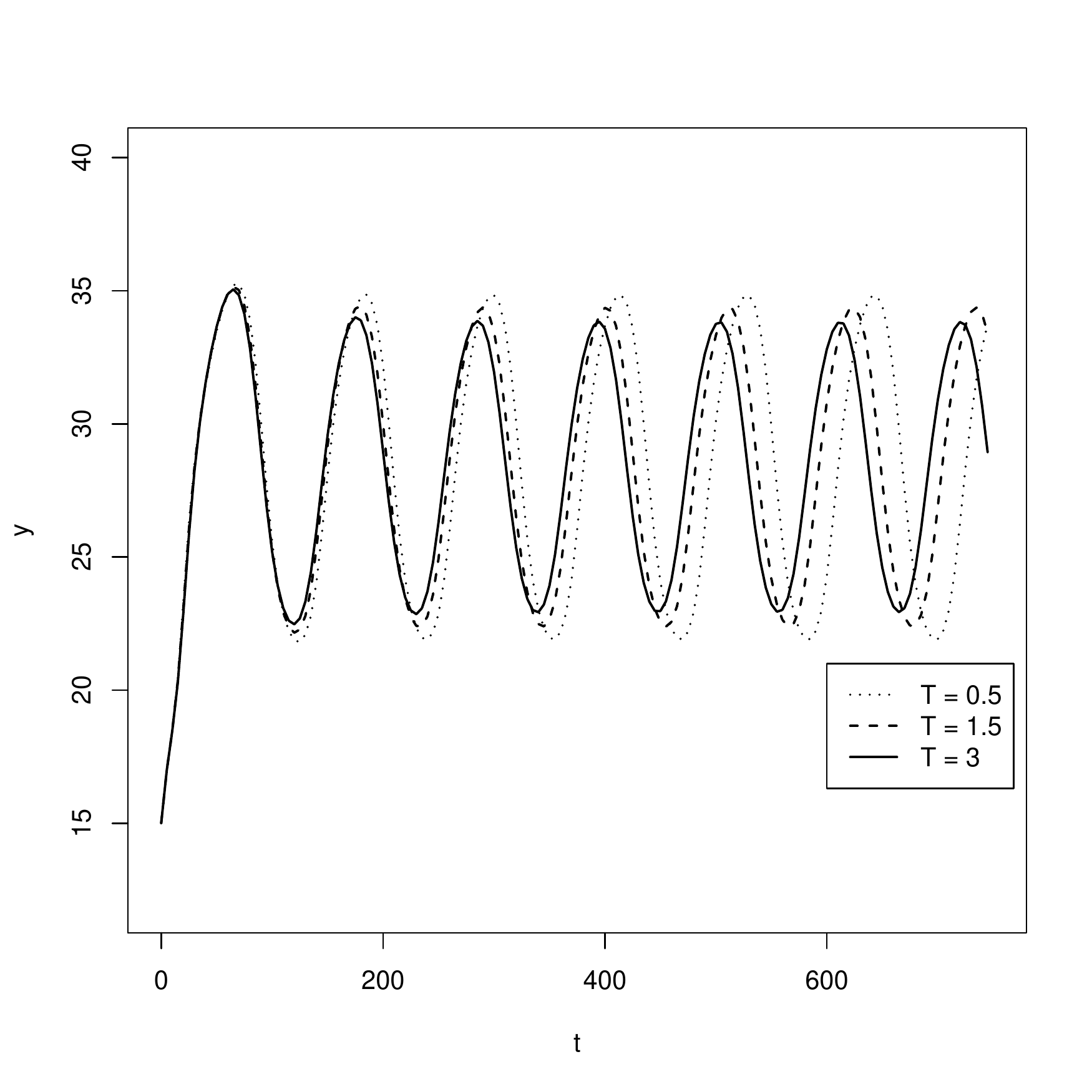}
\end{center}
\caption{\label{fig:6}
Trajectories of model $m=1$ with investment function (\ref{eq:23}) for the parameter $g=0.016$ and $\alpha = 0.6$ (left panel) and $\alpha = 0.9$ (right panel).}
\end{figure}

Now, consider the time paths of the model for the economic variable $y$ for different values of the time delay parameter for the amplitude and period of cycles. In region I there is a stable equilibrium reached by trajectories in an oscillating manner. It is the region of asymptotic stability of the model. In fig.~\ref{fig:6} there are the two cases for $\alpha = 0.6, 0.9$ with three solutions $y(t)$ obtained for given $T = 0.5, 1.5, 3$ and the same initial function $y(t)=1$, $k(t)=100$ where $t \in (-T,0)$. We can see that the dumping of oscillations is weaker as the time delay $T$ increases.

\section*{Case $m=2$}

In this case we consider the four-dimensional model (\ref{eq:10})-(\ref{eq:13}) for the state variables $(y, p, w, k)$. In this model the critical point values of $(y^* = p^* = w^*, k^*)$ are the same as the critical point values of $(y^* = u^*, k^*)$ of three dimensional model presented in the previous section.

\begin{figure}[ht]
\begin{center}
\includegraphics[width=0.45\textwidth]{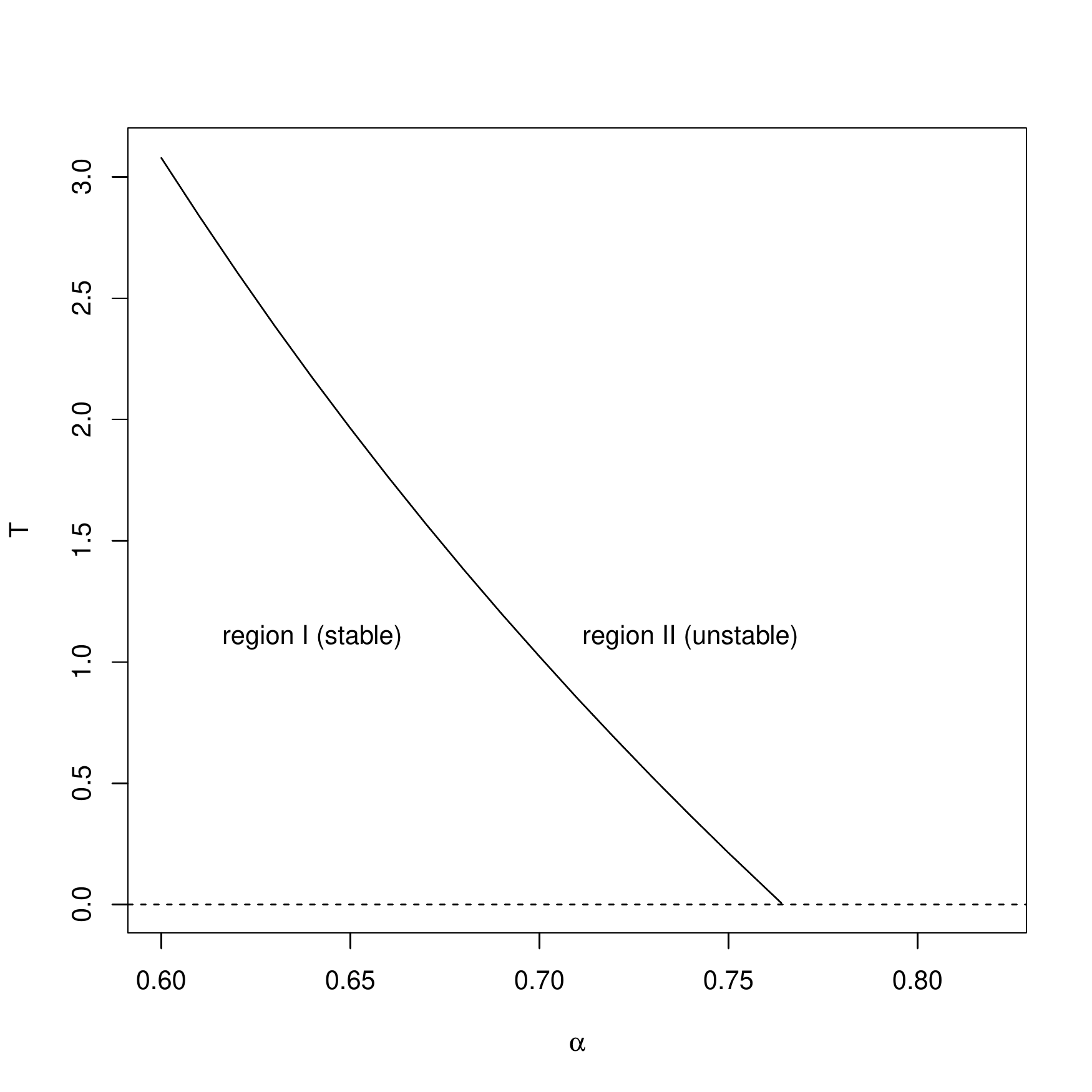}
\end{center}
\caption{\label{fig:7}
The plane of parameters $(\alpha,T)$ for system $m=2$ and parameter $g=0.016$ with investment function (\ref{eq:23}). Region I is the region of asymptotic stability, while region II and III are regions of parameters value for which a limit cycle solution exists.}
\end{figure}

In a similar way as in the previous section we analyse the occurrence of the Hopf bifurcation for the parameter $T$ depending on parameters $\alpha$ and $g$. The relation of $T$ on $\alpha$ for $g=0.016$ is shown in Fig.~\ref{fig:7}. The asymptotic stability region exist only if $\alpha < 0.7644$ (with $g=0.016$). It is the same result as in the case of model $m=1$.

\begin{figure}[ht]
\begin{center}
\includegraphics[width=0.48\textwidth]{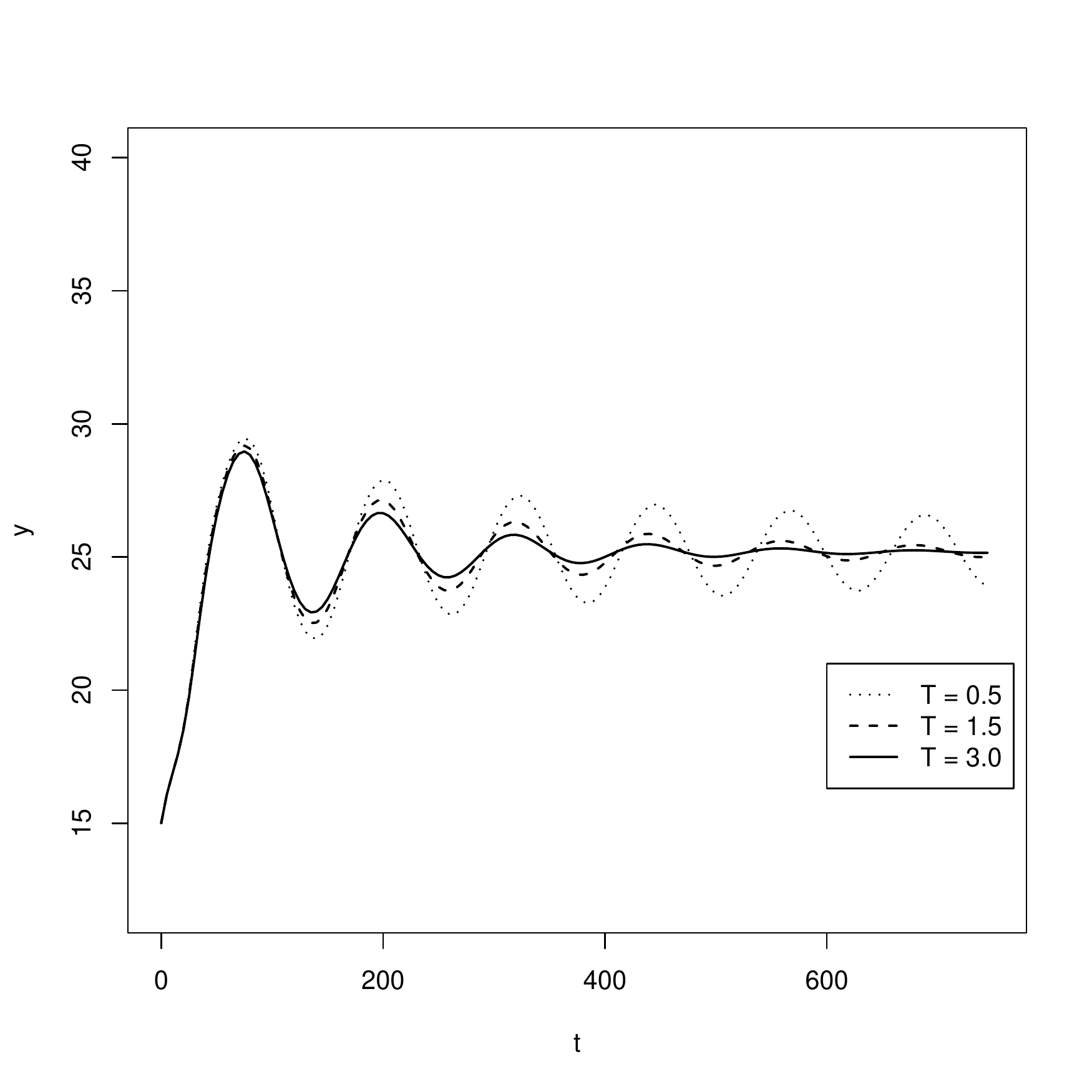}
\includegraphics[width=0.48\textwidth]{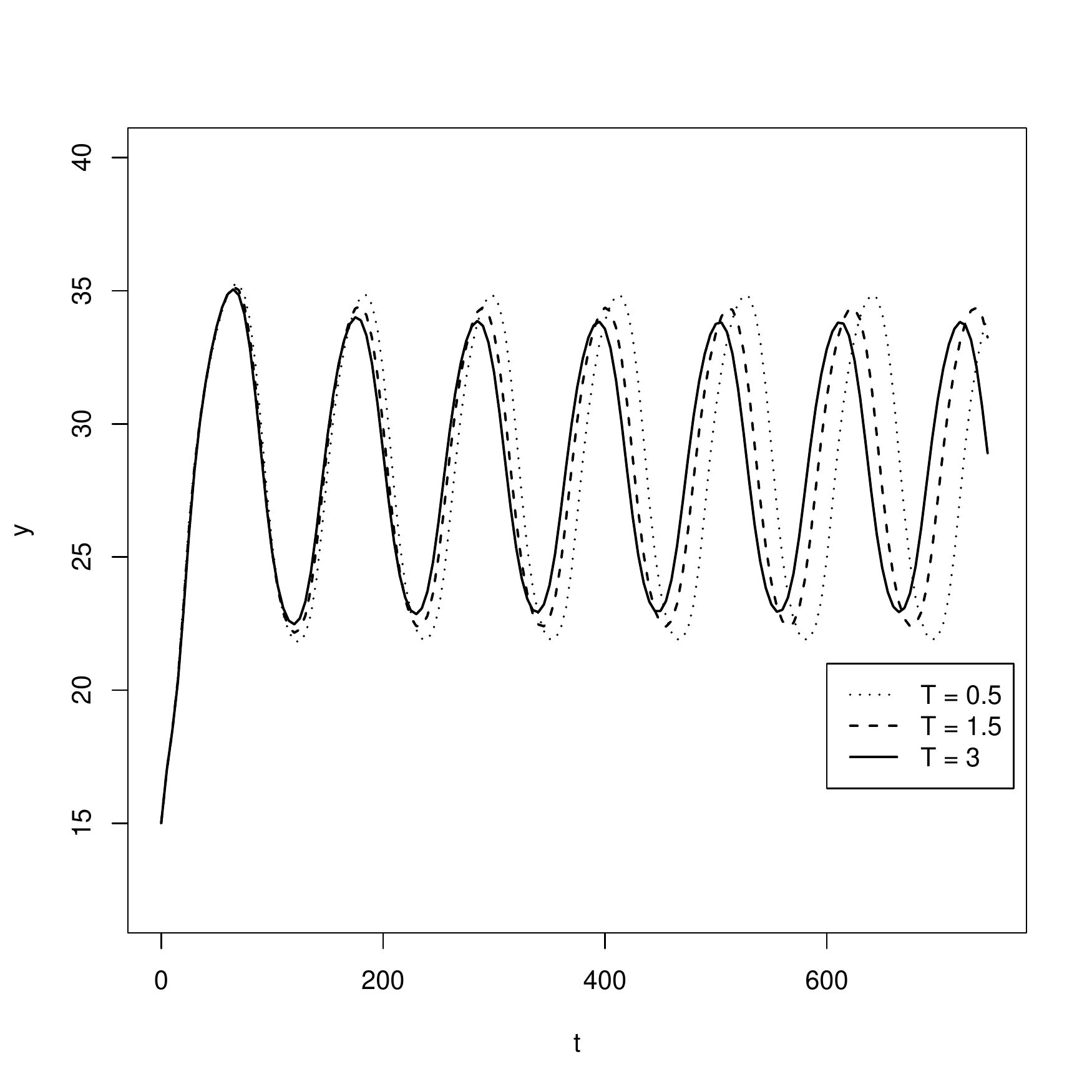}
\end{center}
\caption{\label{fig:8}
Trajectories of model $m=2$  with investment function (\ref{eq:23}) for parameter $g=0.016$ and $\alpha = 0.6$ (left panel) and $\alpha = 0.9$ (right panel).}
\end{figure}

Let us study the cycle characteristics for some values of the parameter $\alpha$ with different values of the delay parameter $T$. Figure~\ref{fig:8} shows the solutions of $y$ for two cases of $\alpha = 0.6, 0.9$ and three values of the parameter $T = 0.5, 1.5, 3.0$. The amplitude and period of cycles are decreasing as the parameter $T$ increases for $\alpha = 0.6$.  

\section*{Comparing $m=1$ and $m=2$}

The models considered can be treated as the approximation of the delay Kaldor-Kalecki growth model. It is important to find how good the subsequent approximations are. Therefore, we compare the bifurcation values of the parameter $T$ in models $m=1$ and $m=2$.

\begin{figure}[ht]
\begin{center}
\includegraphics[width=0.48\textwidth]{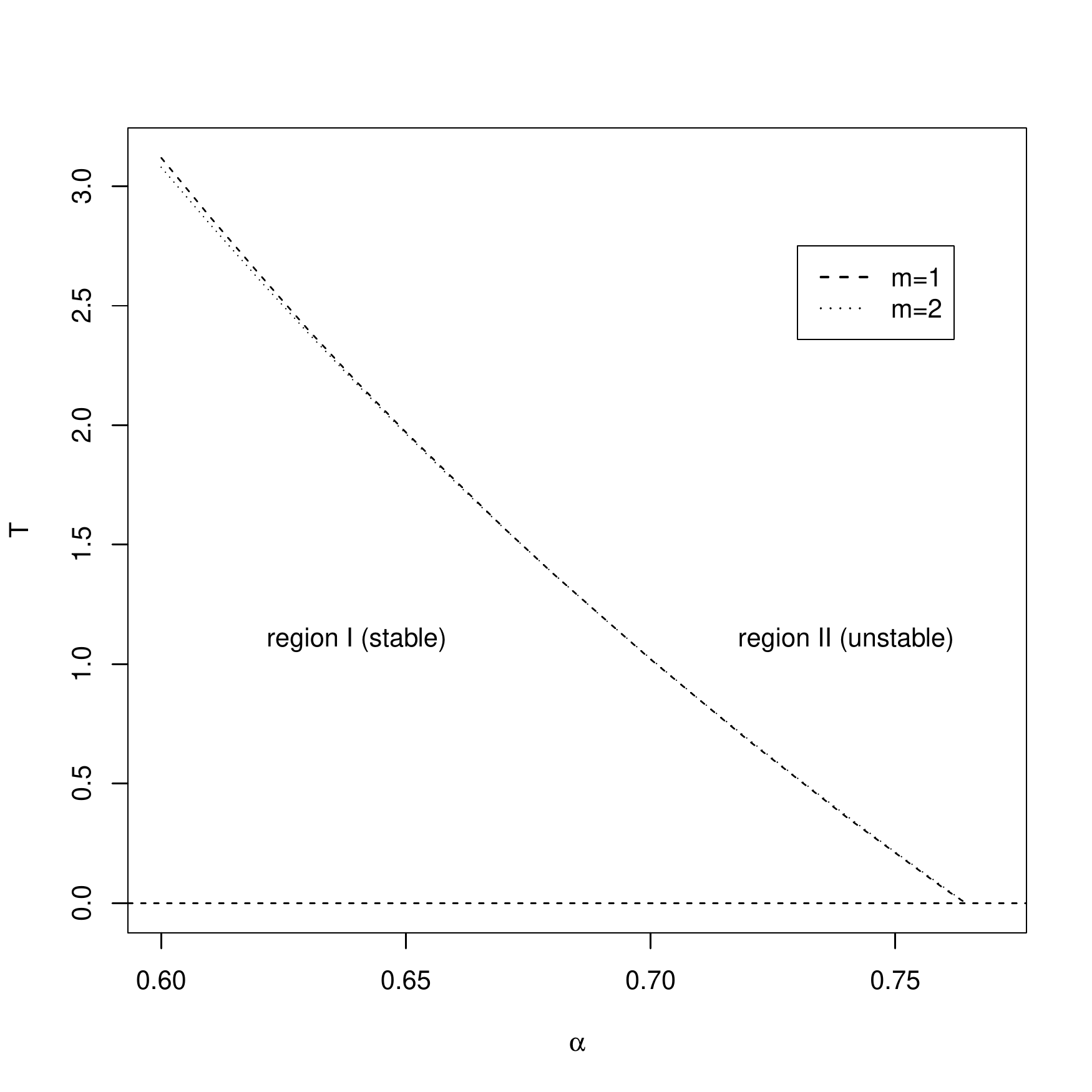}
\includegraphics[width=0.48\textwidth]{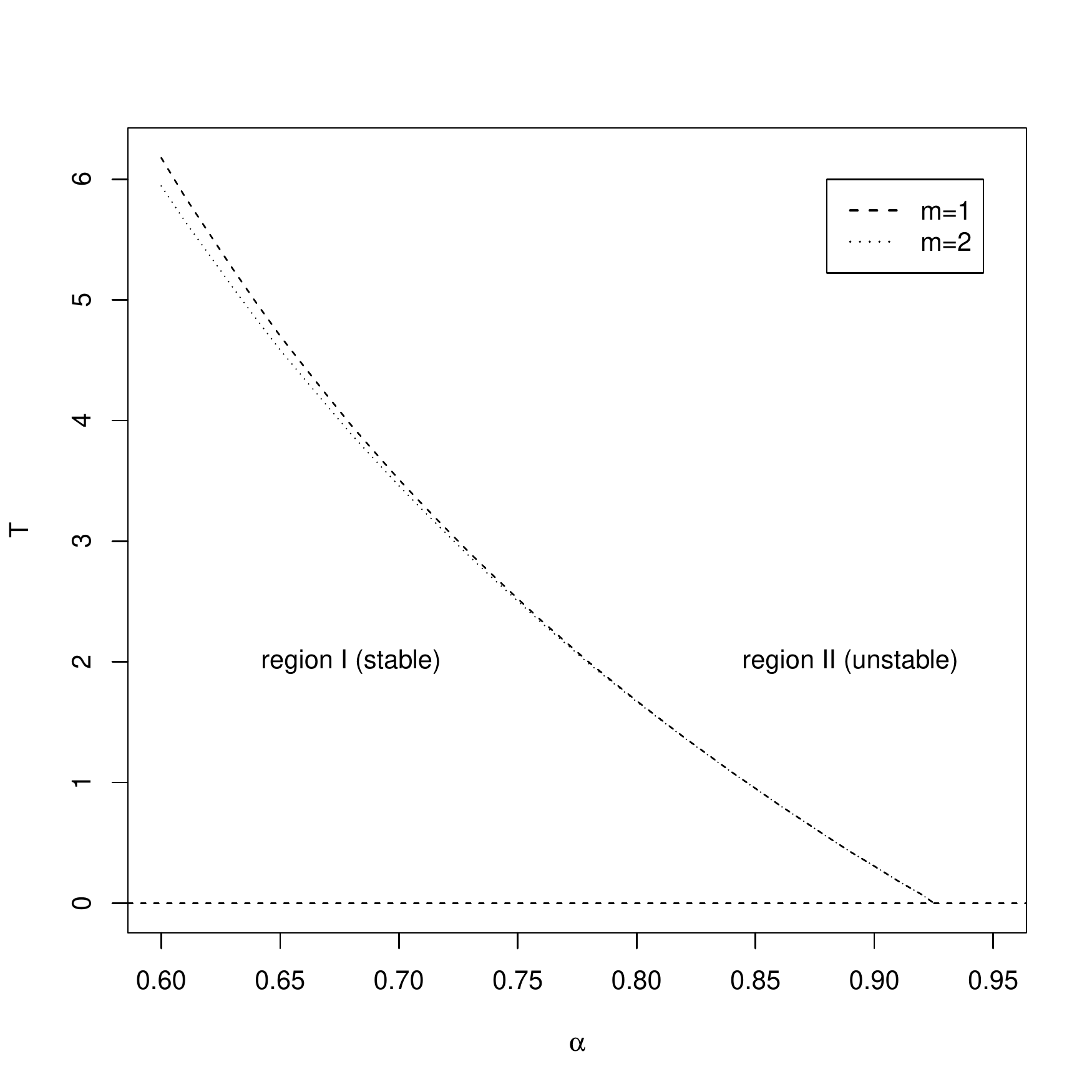}
\end{center}
\caption{\label{fig:9}
The plane of parameters $(\alpha,T)$ for system ($m=1$) and ($m=2$) with investment function (\ref{eq:23}). Here, it is assumed that $g=0.016$ (left panel) and $g=0.011$ (right panel). The dashed line is for model $m=1$ and the dotted line is for model $m=2$. These bifurcation curves separates Region I of asymptotic stability on the left side of curves and region II of limit cycle solution on the right side of curves.}
\end{figure}

First, we consider the bifurcation diagram in the parameter plane $(\alpha, T)$ presented in Fig.~\ref{fig:9}. We find that for the given value of parameter $\alpha$ the Hopf bifurcation value of the parameter $T_{\text{bi}}$ is lower for the model $m=2$. This difference is zero for $T=0$ and then increases as the value $T_{\text{bi}}$ increases for the fixed value of the parameter $\alpha$. On the other hand, for the fixed value of the parameter $T$, the bifurcation value of the parameter $\alpha_\text{bi}$ is greater in the model $m=2$. For the parameter $g=0.016$ the difference is close zero at $\alpha = 0.7644$ and is equal 0.04 at $\alpha = 0.6$ while for the parameter $g=0.011$ the difference is $0.234$ at $\alpha = 0.6$. It is demonstrated in Fig.~\ref{fig:9} for $g=0.016$ (left panel) and $g=0.011$ (right panel).

\begin{figure}[ht]
\begin{center}
\includegraphics[width=0.45\textwidth]{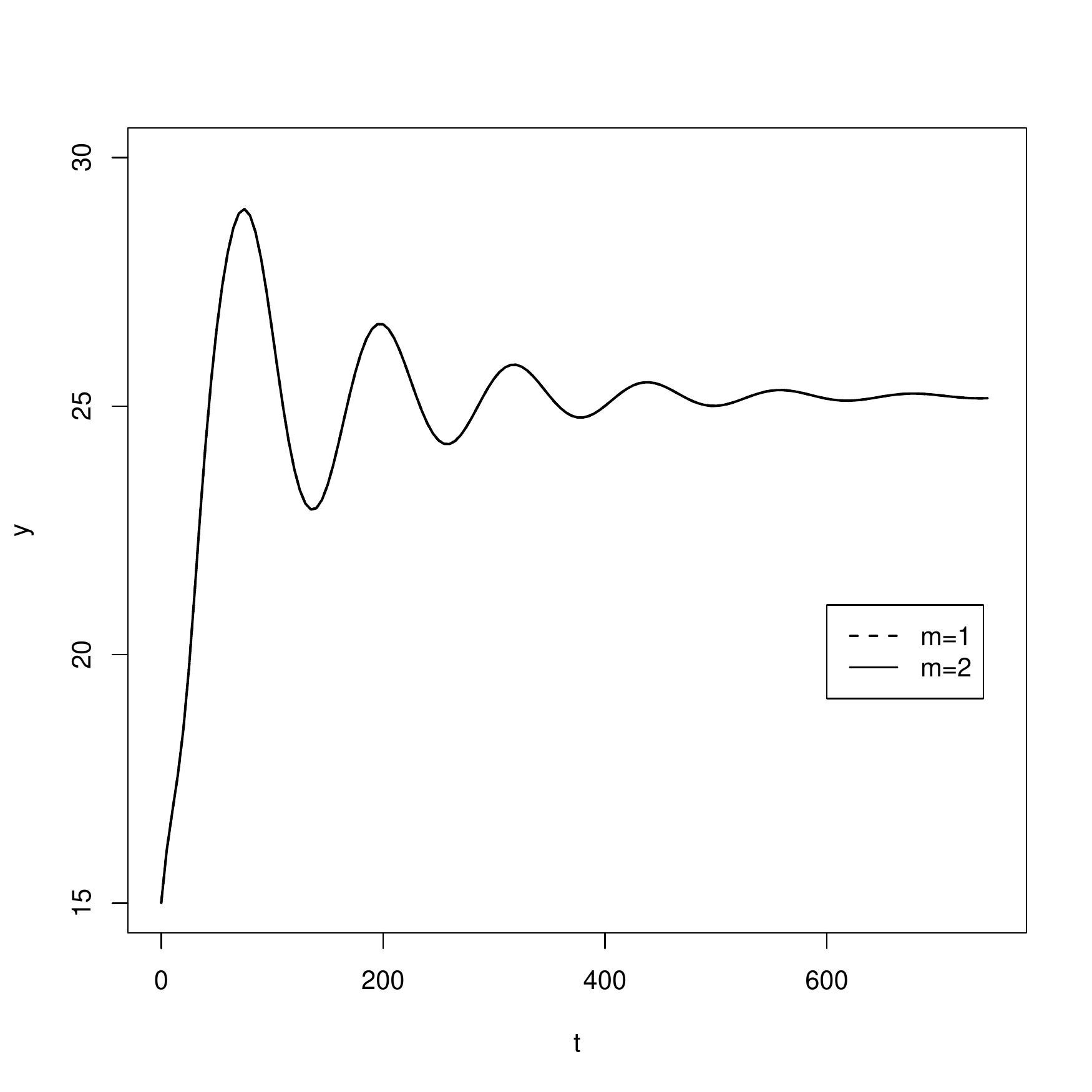}
\includegraphics[width=0.45\textwidth]{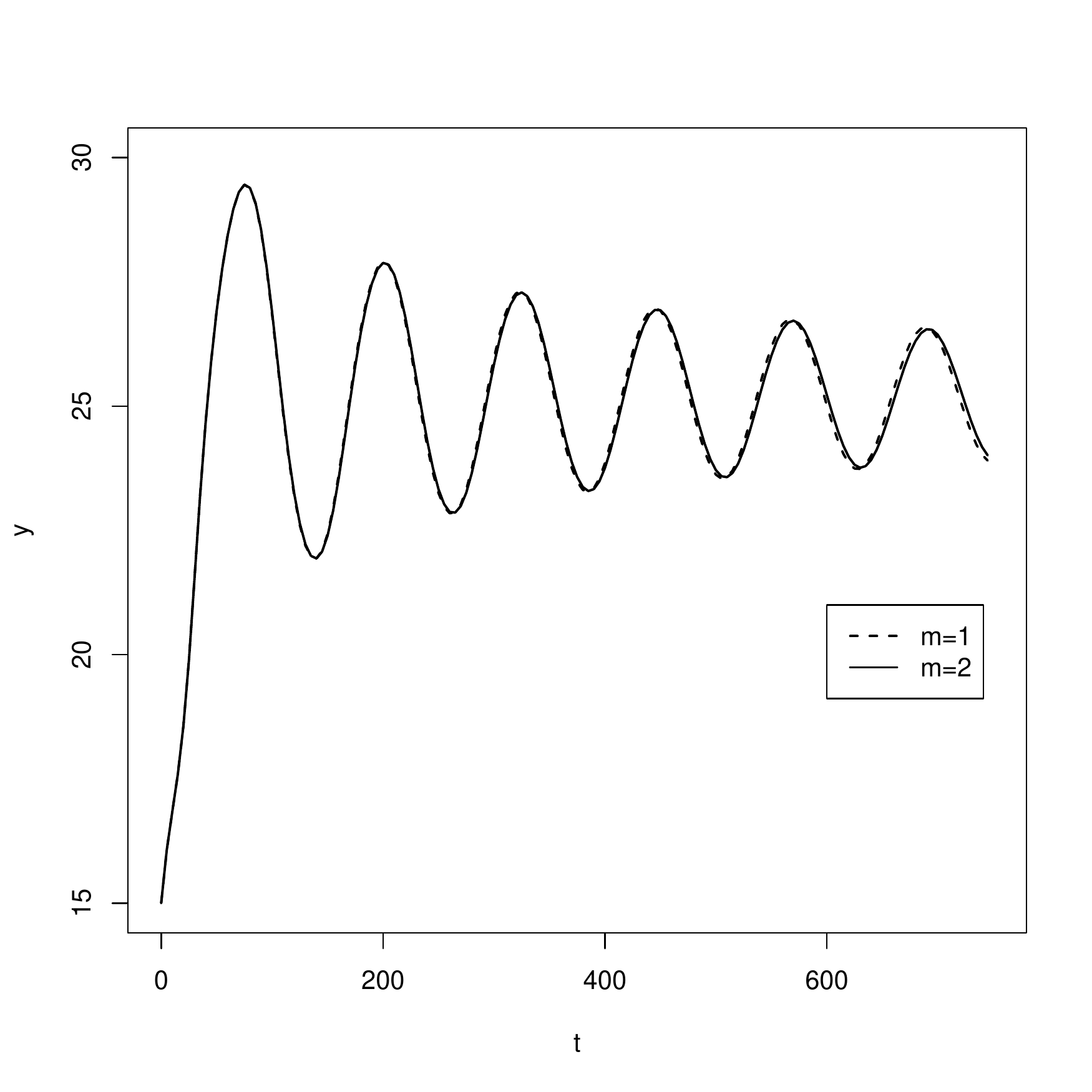}
\end{center}
\caption{\label{fig:10}
Trajectories of models $m=1$ and $m=2$ with investment function (\ref{eq:23}) for the same initial condition $(y=p=w=15, k=100)$ (left panel) for $\alpha =0.6$ and $g=0.016$. The left panel for $T=0.5$ and the right panel for $T=3$.}
\end{figure}

\begin{figure}[ht]
\begin{center}
\includegraphics[width=0.45\textwidth]{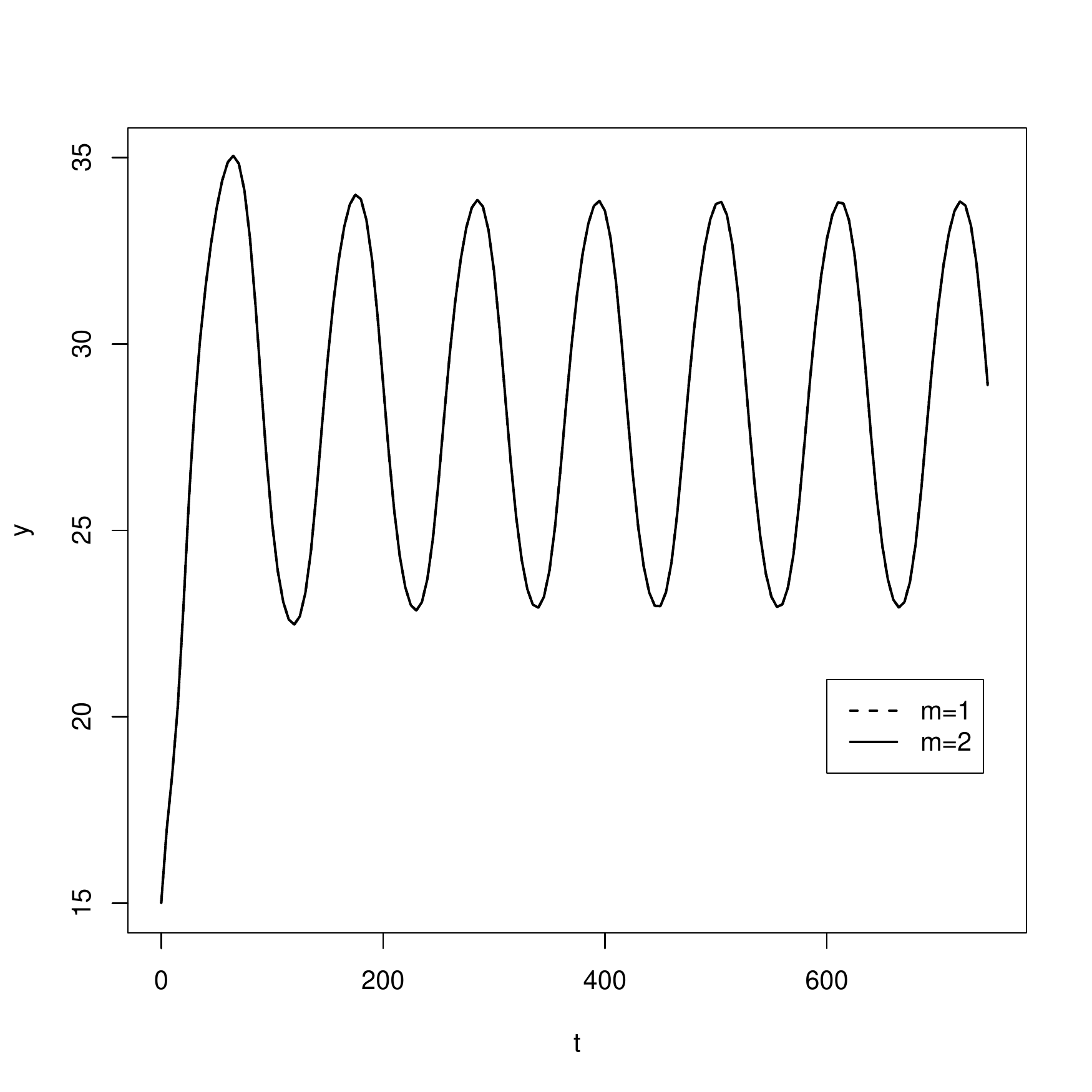} 
\includegraphics[width=0.45\textwidth]{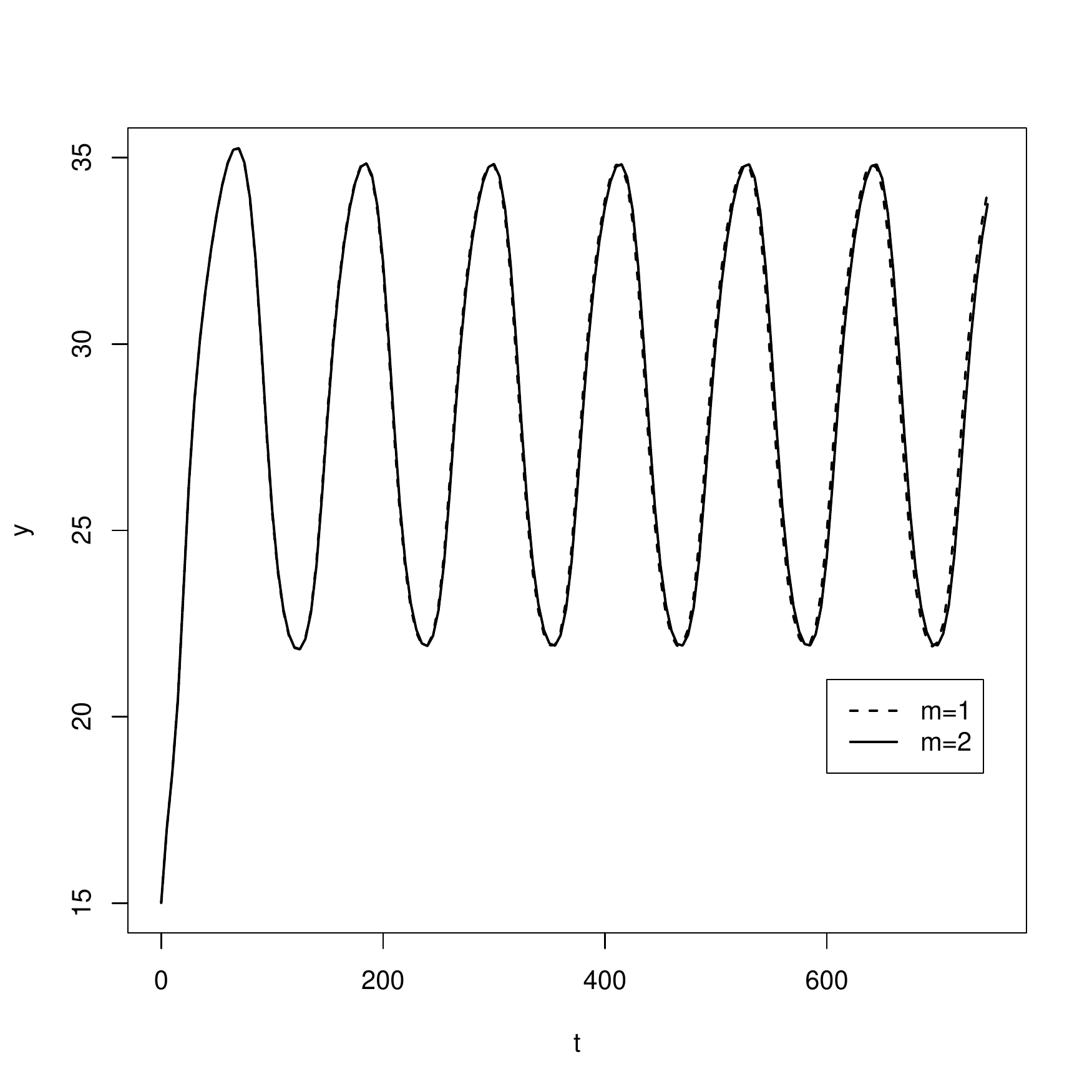} 
\end{center}
\caption{\label{fig:11}
Trajectories of models $m=1$ and $m=2$ with investment function (\ref{eq:23}) for the same initial condition $(y=p=w=15, k=100)$ (left panel) for $\alpha = 0.9$ and $g=0.016$. The left panel for $T=0.5$ and right panel for $T=3$.}
\end{figure}

We compare trajectories of $y(t)$ for systems $m=1$ and $m=2$ with the same initial conditions. In Figs.~\ref{fig:10}, \ref{fig:11} there are trajectories $y(t)$ for assumed the parameter $g=0.016$ and combinations of parameters $\alpha$ and $T$. We find that for the same parameters $\alpha$, $g$ and $T$ the period of cycles is smaller for model $m=2$ and amplitude is also smaller, although the difference is very small. For example, for $\alpha = 0.9$, $g=0.016$ and $T=3$, the period of cycle in model $m=1$ is $114.85$ and in model $m=2$ is $116.45$ while amplitudes are $12.9555$ and $12.966$, respectively. Taking models with greater $m$ we should obtain the cycles with longer periods and amplitudes.

We can explore further approximations of the model and compare values of the bifurcation parameter $g$ for successive $m$ in table~\ref{tab:2}.

\begin{table}
\begin{center}
\caption{The Hopf bifurcation points $g_\text{bi,1}$ and $g_\text{bi,2}$ for subsequent approximations}
\label{tab:2}
\begin{tabular}{ccc}
\toprule
model & $g_\text{bi,1}=1$ & $g_\text{bi,2}=2$ \\
\midrule
$m_{1}$ & 0.01011989 & 0.02032586 \\
$m_{2}$ & 0.01011919 & 0.02032671 \\
$m_{3}$ & 0.01011909 & 0.02032693\\
$m_{4}$ & 0.01011906 & 0.02032703 \\
\bottomrule
\end{tabular}
\end{center}
\end{table}

\section{Conclusions}

We study the Kaldor-Kalecki growth model with distributed delay transform to the ordinary differential equation system using the linear chain trick. It allows to choose the arbitrary dimension of the system. We consider two cases where the system is transformed to three-dimensional and four-dimensional dynamical systems.

The model possesses the equilibrium for some interval of values of rate of growth parameter. The interval depends on the values of investment function which is taken to model analysis.

We study a bifurcation to a limit cycle (the Hopf bifurcation) due to the change of two parameters: the time delay $T$ and the rate of growth $g$. Additionally we take take into consideration the speed of adjustment parameter $\alpha$ which is the bifurcation parameter of original Kaldor model. For the better insight of dynamics we analyzed the bifurcations in the the three dimensional space of these parameters.

When we consider the dynamics of the model under the change of the growth rate parameter, we discover numerically two bifurcation values of the rate of growth parameter when the Hopf bifurcations occur. Increasing the value of the rate of growth parameter, for a smaller value of this parameter the limit cycle emerges then for a larger value of this parameter the limit cycle is destroyed to a stable focus through the Hopf bifurcations. Therefore, the cyclic behaviour takes place in some interval of the rate of growth parameter values. Outside of the interval the systems through damping oscillation goes to a stable stationary solution. The similar dynamic behaviour with two Hopf bifurcations separating stable, unstable and stable regions was also found in the macroeconomic model extending the Calvo and Obstfeld framework \cite{Mavi:201970}.

There are two oscillating regimes. For lower and higher rates of growth the oscillations are damped and asymptotically stationary state is reached. For intermediate rates of growth the self-sustained oscillations of constant amplitude are present. For some model parameters this intermediate interval of rate of growth values is obtained to be $(0.01011989,0.0238466)$ for 3-dimensional and 4-dimensional systems. The range of this intermediate interval depends on the model parameters: $\alpha$, $\gamma$ and $\delta$.

All numerical analyses have been done with Dana and Malgrange's investment function for the French macroeconomic data \cite{Dana:1984dd}.

\begin{itemize}
\item The Kaldor-Kalecki model with distributed delay is reduced to the ordinary differential system using the linear chain trick technique.
\item Depending on the value of the parameter $m$ of the $\Gamma$ distribution function the reduced system is $(m+2)$-dimensional ordinary differential equation system.
\item For the increasing time delay parameter there is the supercritical Hopf bifurcation.
\item For the increasing rate of growth parameter, first the limit cycle emerges and then the limit cycle disappears. Therefore, there are two supercritical Hopf bifurcations with two bifurcation values of the rate of growth parameter.
\item For some values of parameters $\alpha$ and $T$, in the allowed range of the rate of growth parameter values, both for lower and higher values of the rate growth parameter the model has the stable stationary point while for the middle range of parameter values there is the limit cycle.
\item The period of cycle increases and decreases as the rate of growth parameter increases in the range of unstable solution.
\item Comparing the models with different $m$ the stable region in the parameter space is slightly diminished as $m$ is greater.
\end{itemize}

\subsection*{Acknowledgements}
A.K. and M.S. acknowledge the support of the Narodowe Centrum Nauki (Polish National Science Centre) project 2014/15/B/HS4/04264.

\end{document}